\documentclass[twocolumn]{aastex63}

\usepackage{graphicx}   
\usepackage{subfigure}
\usepackage{multirow}
\usepackage{amsmath}

\graphicspath{{./}{figures/}}
\shorttitle{profiles of filaments since z=4.0}
\shortauthors{Zhu, Zhang \& Feng.}

\begin{document}

\title{Profiles of cosmic filaments since z=4.0 in cosmological hydrodynamical simulation}

\correspondingauthor{Weishan Zhu}
\email{zhuwshan5@mail.sysu.edu.cn}

\author[0000-0002-1189-2855]{Weishan, Zhu}
\affil{School of Physics and Astronomy, Sun Yat-Sen University, Zhuhai campus, No. 2, Daxue Road \\
Zhuhai, Guangdong, 519082, China}

\author{Fupeng, Zhang}
\affil{School of Physics and Materials Science, Guangzhou University\\
Guangzhou, Guangdong, 510006, China}

\author{Long-Long, Feng}
\affil{School of Physics and Astronomy, Sun Yat-Sen University, Zhuhai campus, No. 2, Daxue Road \\
Zhuhai, Guangdong, 519082, China}



\begin{abstract}
A large portion of the baryons at low redshifts are still missing from detection. Most of the missing baryons are believed to reside in large scale cosmic filaments. Understanding the distribution of baryons in filaments is crucial for the search for missing baryons. We investigate the properties of cosmic filaments since $z=4.0$ in a cosmological hydrodynamic simulation, focusing on the density and temperature profiles perpendicular to the filament spines. Our quantitative evaluation confirm the rapid growth of thick and prominent filaments after $z=2$. We find that the local linear density of filaments shows correlation with the local diameter since $z=4.0$. The averaged density profiles of both dark matter and baryonic gas in filaments of different width show self-similarity, and can be described by an isothermal single-beta model. The typical gas temperature increases as the filament width increasing, and is hotter than $10^6$ K for filaments with width $D_{fil} \gtrsim 4.0 \rm{Mpc}$, which would be the optimal targets for the search of missing baryons via thermal Sunyaev-Zel’dovich (SZ) effect. The temperature rises significantly from the boundary to the inner core regime in filaments with $D_{fil} \gtrsim 4.0 \rm{Mpc}$, probably due to heating by accretion shock, while the temperature rise gently in filaments with  $D_{fil}< 4.0 \rm{Mpc}$.

\end{abstract}



\section{Introduction} \label{sec:intro}

Baryons are expected to comprise $\sim 5 \%$ of the energy density in the universe according to the concordance $\Lambda \rm{CDM}$ cosmology and observations such as the cosmic microwave background(e.g. \citealt{2014A&A...571A..16P}). The abundance of baryonic matter at high redshifts, revealed by the lyman-$\alpha$ forest, agrees with the prediction of $\Lambda \rm{CDM}$ cosmology(\citealt{1997ApJ...489....7R};\ \citealt{1997ApJ...490..564W}). However, the baryons that have been detected at redshift $z<2$ fall short of the expectation from the standard $\Lambda \rm{CDM}$ model by a substantial portion. At low redshifts, around $10 \%$ of the baryons reside in galaxies as stars and interstellar medium, and another $10 \%$ are found to be circumgalactic medium and diffuse medium in galaxy groups and clusters(\citealt{1998ApJ...503..518F};\ \citealt{2012ApJ...759...23S}). The remaining baryons are expected to reside outside of collapsed halos. About $25\%-42 \%$ of the baryons in relatively cool state($T \sim 10^4 \rm{K}$) have been detected by the Lyman-alpha forest and broad Lyman-alpha absorbers(BLA), and $\sim 11\%$ have been observed by extragalactic OVI absorbers(\citealt{2012ApJ...759...23S};\ \citealt{2016ApJ...817..111D}; \ \citealt{2016MNRAS.455.2662T};). Note that, these fractions associated to HI and OVI absorbers may have considerable uncertainties(e.g.\citealt{2020arXiv201209203T}).

By and large, $30-50\%$ of the baryonic matter are "missing" from detection at low redshifts. On the other hand, cosmological simulations suggest that a significant fraction of the baryonic matter is residing in filamentary and sheet like structures of the cosmic web ( \citealt{1999ApJ...514....1C};\ \citealt{2001ApJ...552..473D};\ \citealt{2006MNRAS.370..656D};\ \citealt{2016MNRAS.457.3024H}; \ \citealt{2017ApJ...838...21Z}; \ \citealt{2018MNRAS.473...68C}; \ \citealt{2019MNRAS.486.3766M}). Those baryons residing in filaments and sheets are predicted to be 'warm-hot' with temperatures $10^5-10^7 \rm{K}$, and over-densities of $\delta_b \sim 0-100$. This diffuse medium is named as warm-hot intergalactic medium(WHIM). In the past two decades, many efforts have been made to detect this WHIM via different observational tools. Actually, those baryons detected via extragalactic OVI absorbers(e.g. \citealt{2016ApJ...817..111D}) should be WHIM with temperature $10^5-10^{5.5} \rm{K}$. However, it is very challenge to observe the  WHIM in hotter phases with $T=10^{5.5}-10^7 \rm{K}$. Once most of the WHIM are located, it would largely solves the "missing baryon" problem.

Many efforts have been paid to observe the WHIM via X-ray emission. However, as the emissions are quite faint, currently there are only a limited number of observations that mainly focus on a few individual filaments linking or around massive clusters(\citealt{2007ARA&A..45..221B};\ \citealt{2009ApJ...699.1765B}; \ \citealt{2015Natur.528..105E}; \citealt{2017A&A...606A...1A} ). Very recently, \citealt{2020A&A...643L...2T} reports a $4.2\sigma$ detection of stacked X-ray emission from the WHIM in filaments. Meanwhile, possible X-ray absorption by ions in the WHIM toward background quasars has also been reported in the literatures(e.g. \citealt{2002ApJ...572L.127F};\ \citealt{2005ApJ...629..700N};\ \citealt{2016MNRAS.457.4236B};\ \citealt{2018Natur.558..406N};\ \citealt{2019A&A...621A..88N}). However, those studies are not statistically significant, and some of them lack further confirmation.

Given the current challenge to locate the WHIM with X-ray emission and absorption, alternative tools, including the Sunyaev-Zel'dovich (SZ) effect, have been proposed. Several recent works have claimed the detection of thermal SZ signal due to the WHIM in filaments between galaxy pairs, or galaxy clusters(\citealt{2018A&A...609A..49B};\ \citealt{2019MNRAS.483..223T};\ \citealt{2019A&A...624A..48D}; \ \citealt{2020A&A...637A..41T}). These works demonstrate that the SZ effect could be a powerful tool to locate the WHIM. Meanwhile, interpretation of these signals involves the knowledge of gas density and temperature in filaments. The assumed or estimated density and temperature profiles have notable differences between these works. A well understanding of the properties of baryonic matter in filaments is desired to justify the recent observational results and help to locate the WHIM with current and future detection via SZ effect and X-ray emissions and absorption.

Recently, \cite{2019MNRAS.486..981G}, \cite{2020arXiv201015139G} and \cite{2020arXiv201209203T} have investigated the distribution of gas in filaments based on three different cosmological simulations. The filaments in different simulation samples are all found to have isothermal cores, despite the numerical methods used to identify filaments are different. However, there are notable differences on the central temperature, core radius, and mass density profiles. Moreover, the density and temperature profiles measured in simulation samples also show notable differences with the estimations in those observational works that reporting the detection of tSZ signals from filaments. It is worthwhile to probe the profiles of filaments in more simulation samples generated with different codes, and with different identification method of filaments, as well as with different stacking/averaging methods for the profile measurement. 

For instance, filaments with similar length or overdensity are stacked to obtain the average profiles in the literature. Few works have try to measure the density and temperature profiles of gas in filaments with similar width/thickness. \citealt{2014MNRAS.441.2923C} demonstrate that the width is a better indicator than length to describe the evolution of prominent filaments, which may contain more hot IGM, i.e more likely detected via SZ effect and X-ray, than tenuous filaments. In addition, the evolution of filament profiles with redshift is barely known. 

In this work, we make use of samples from a cosmological hydrodynamic simulation with adaptive mesh refinement to study the evolution of cosmic filaments, focusing on the density and thermal profiles. We measure the profiles of filaments with similar width. This paper is organised as follows: Section 2 introduces the cosmological simulation used here and the numerical methods employed for classifying filaments and measuring their diameter, density and thermal profiles. Section 3 presents the number frequency, linear density, measured density and temperature profiles of filaments and their evolution since redshift $z=4$. In section 4, we assess the impact of threshold values used in cosmic web classification, and compare our results with previous simulation and observational studies. Finally, our findings are summarized in Section 5.  

\section{Methodology} \label{sec:method}

\begin{figure*}[htb!]
\begin{center}
\hspace{-0.0cm}
\includegraphics[width=0.48\textwidth, trim=80 40 40 30, clip]{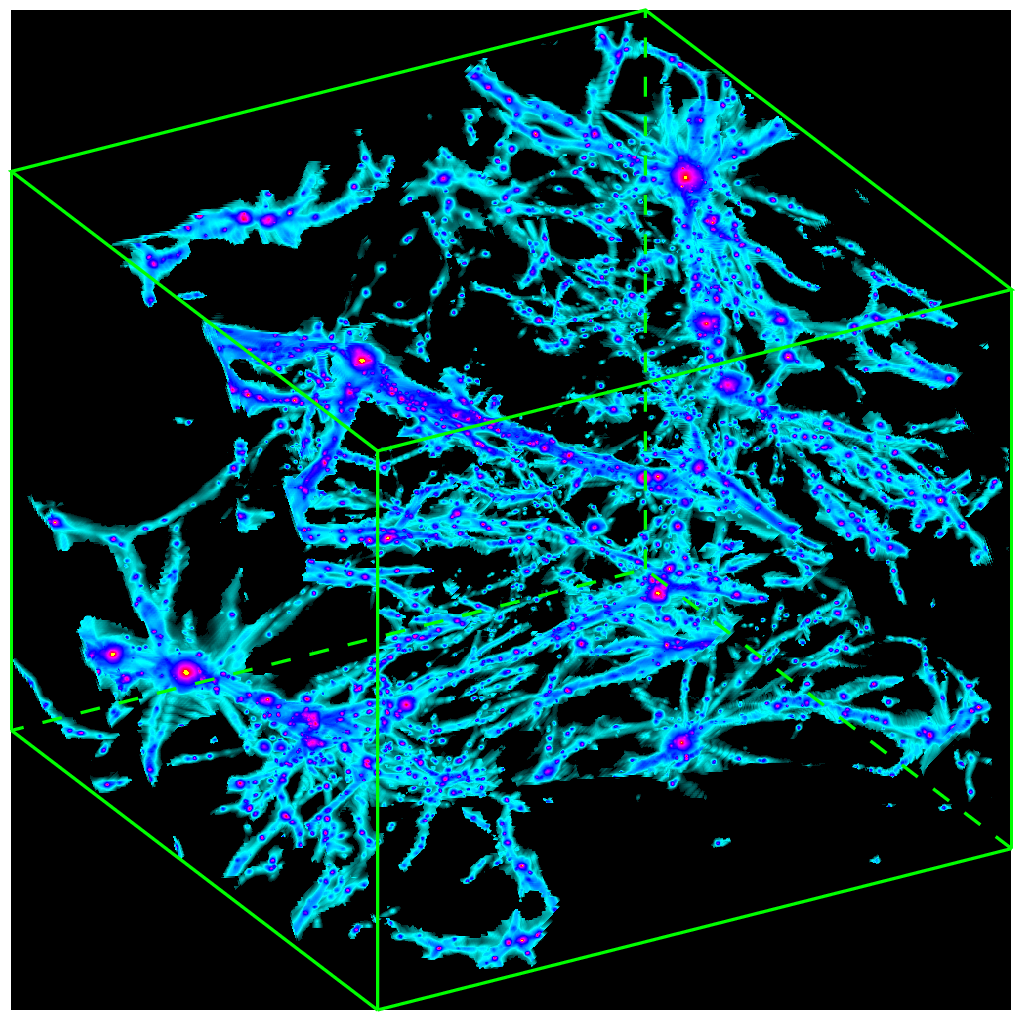}
\includegraphics[width=0.48\textwidth, trim=80 40 40 30, clip]{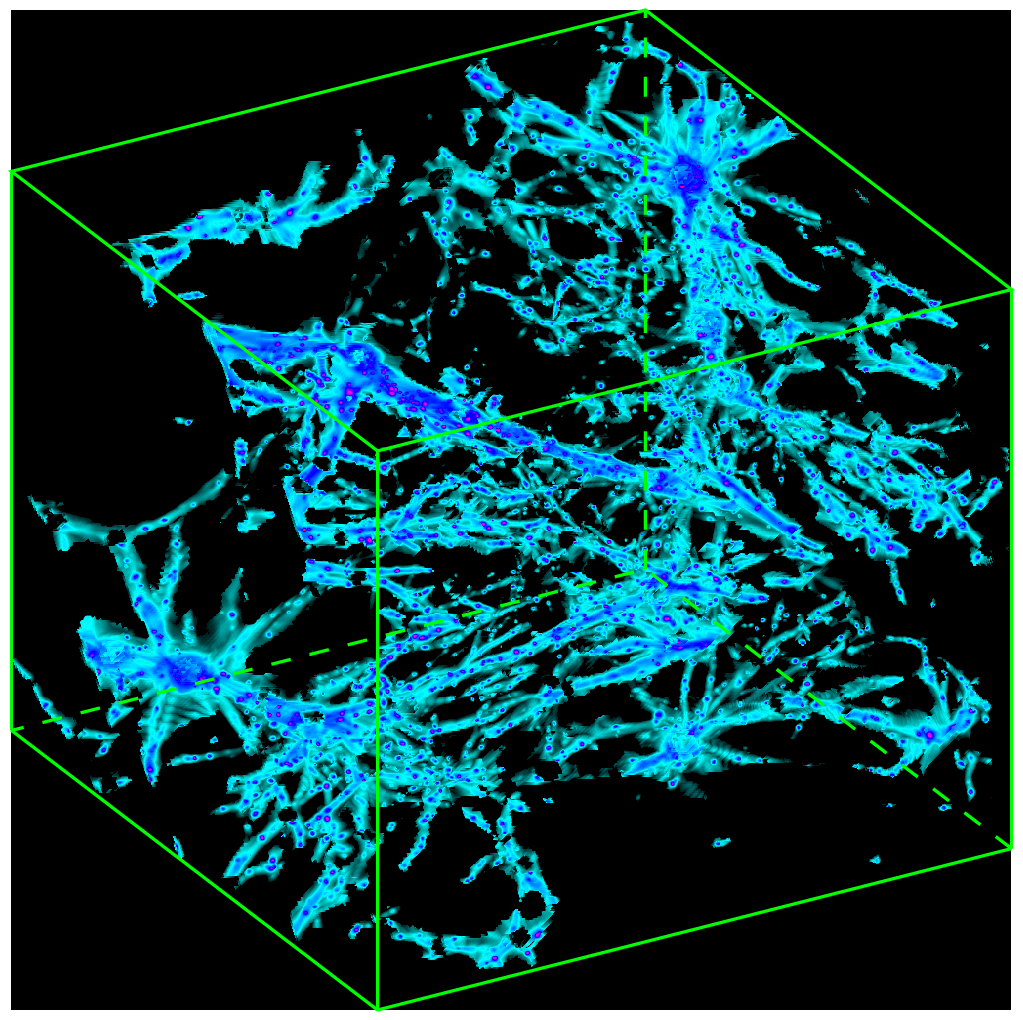}
\includegraphics[width=0.48\textwidth]{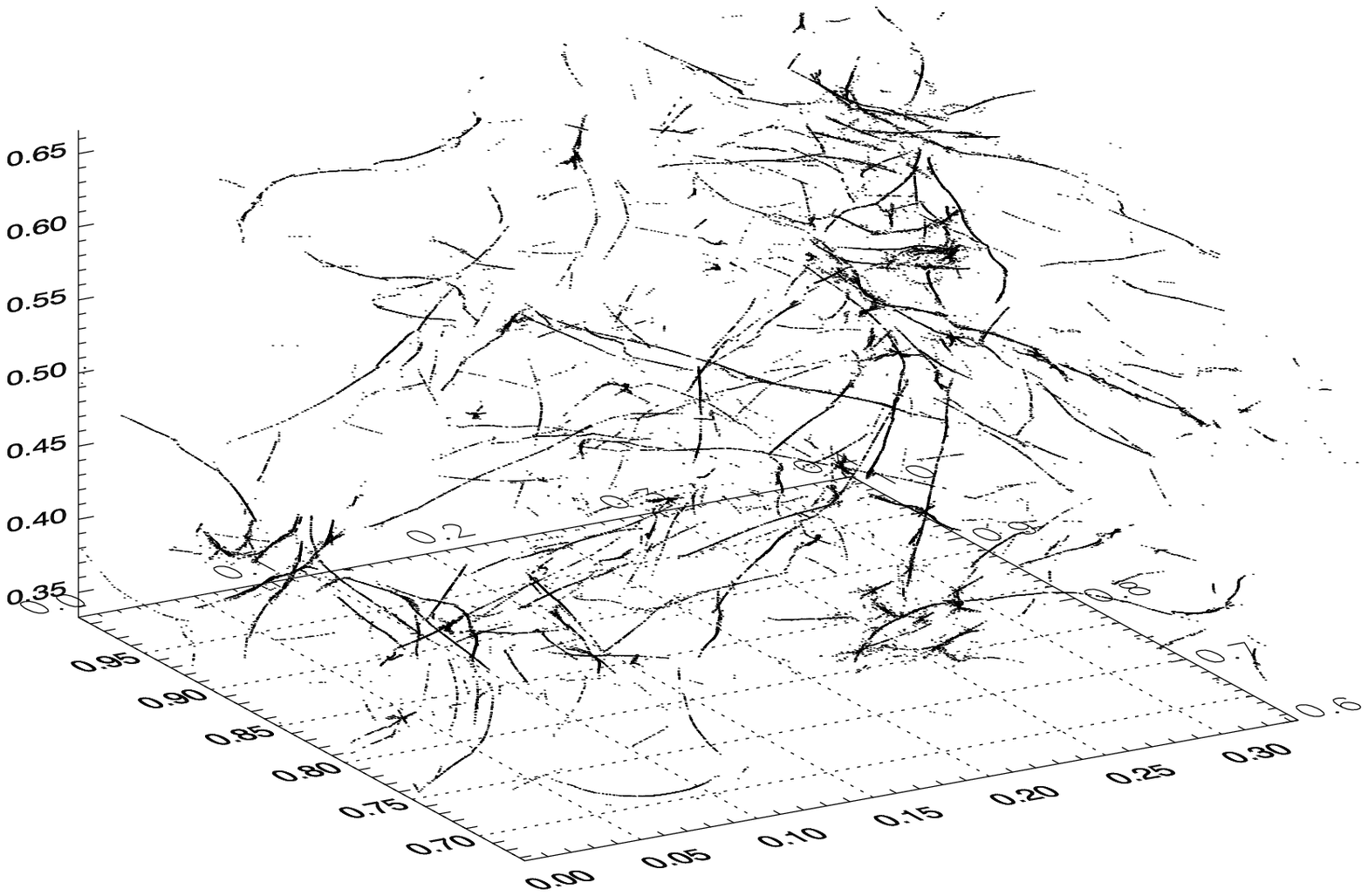}
\includegraphics[width=0.48\textwidth, trim=80 40 40 30, clip]{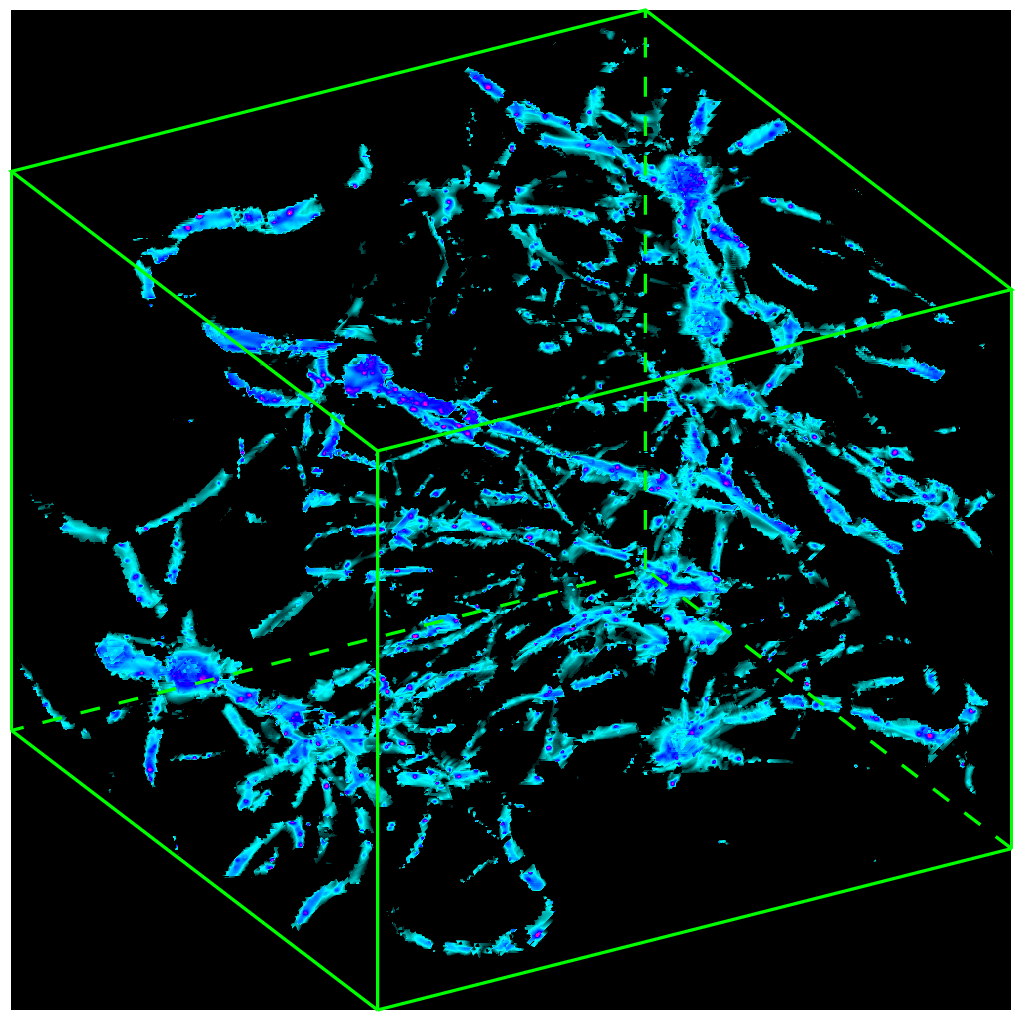}
\caption{Top left: Density field of baryonic matter in grid cells belonging to filaments and clusters in a cubic box of $(33.3 h^{-1}\rm{Mpc})^3$; Top right: the same as the top left panel but present grid cells belonging to filaments only. Bottom left: $5\%$ of the grids belonging to filaments after compressed. Bottom right: As in the top right plot, but only present grids with a distance to the spine that is less than half of the local radius, i.e. $r_{nml}<0.5$. }
\label{fig:fila_comp}
\end{center}
\end{figure*}

\subsection{Simulation Samples} 
We use samples from a cosmological hydrodynamic simulation with adaptive mesh refinement, which is run by the cosmological code RAMSES(\citealt{2002A&A...385..337T}). The simulation has a volume of $(100 h^{-1})^3$ Mpc, assuming a $\Lambda$CDM cosmology with parameters  $\Omega_{m}=0.317, \Omega_{\Lambda}=0.683,h=0.671,\sigma_{8}=0.834, \Omega_{b}=0.049$, and $n_{s}=0.962$(\citealt{2014A&A...571A..16P}). The simulation is evolved with $1024^3$ dark matter particles, and $1024^3$ root grid cell. An AMR grid up to level $l_{max}=17$ is adopted. Namely, the spacial resolution is $97.6 h^{-1}$kpc for the root grid, and reaches $0.763 h^{-1}$kpc at the finest level.  Radiative cooling and heating of gas, star formation and stellar feedback are implemented, while feedback from active galactic nuclei(AGN) is not included.  A uniform UV background assuming the model in \cite{1996ApJ...461...20H} is switched on at redshift $z=8.5$. More details about the simulation can be found in \cite{2021ApJ...906...95Z}. This simulation starts at $z=99$ and ends at $z=0$. In this work, we use samples from z=4.0 to z=0.0. 

\subsection{filaments classification, compress and measurement}
From the simulation sample, we build the density of baryonic and dark matter on a $512^3$ grid respectively. The resampled density fields on the $512^3$ grids are smoothed with a Gaussian kernel of radius $0.39/h$ Mpc. Then we assign the grid cells into four categories of cosmic large scale structures, i.e., clusters/nodes, filaments, sheets/walls and voids for baryonic and dark matter respectively. We apply the tidal tensor, i.e., the Hessian matrix, of the rescaled peculiar gravitational potential, to complete the classification.  For each grid cell, its type of cosmic web is determined by counting the number of eigenvalues larger than a given threshold $\lambda_{th}$. We refer the readers to \cite{2007MNRAS.375..489H}, \cite{2009MNRAS.396.1815F} and \cite{2017ApJ...838...21Z} for more details about this web classification scheme and the choice of $\lambda_{th}$. In this work, we adopt $\lambda_{th}=0.2$, and find that filaments occupy around $14\%$ of the volume, and contains $\sim 45\%$ of the mass at $z=0$. The volume filled by filaments changes slightly between $z=0$ and $z=3$, while the mass fraction declines moderately to $34\%$ at $z=3$. 

We follow the procedures in \cite{2014MNRAS.441.2923C}(see their Section 4 and Appendix) to compress the grids in filaments, find the spines of filaments, and then estimate the contribution to filament length, the local linear density, $\zeta_{fil}$, and the local width/diameter of filaments , $D_{\rm{fil}}$, for each grid in filaments. We further set $R_{fil}=D_{\rm{fil}}/2$ to stand for the local radius of filaments. We use the same spherical filter of radius $R=1 h^{-1}$Mpc as in \cite{2014MNRAS.441.2923C} to search neighbours of grids during the filaments compressing, and take a segment of length $\Delta L= 3 h^{-1}$Mpc for estimating its contribution to filament length and the local width/thickness. Figure \ref{fig:fila_comp} illustrates the effect of the compress procedures. The top left plot in Fig. ~\ref{fig:fila_comp} shows the density field of baryonic matter residing in filaments and clusters/nodes in a subbox of volume $(33.3 h^{-1} \rm{Mpc})^3$, while the top right plot shows filaments only. The position of grid cells in filaments after compress are presented in the bottom left panel. Based on the compressed filaments, we measure the distance of each grid cell in filaments to the spine along the direction perpendicular to the orientation of filament. We denote this distance as $r$, and introduce a rescaled/normalised radial distance to spine for each filament grid defined by $r_{\rm{nml}}=r/R_{\rm{fil}}$. The bottom right panel in Fig. ~\ref{fig:fila_comp} shows the density filed of grid cells in filament with $r_{\rm{nml}}<0.5$, i.e. the relatively inner region of filaments.

\section{Properties of filaments} \label{sec:properties}
We first probe the evolution of filament width, the relation between local width and local linear density, and then investigate the density and temperature profiles of filaments.

\begin{figure*}[htbp]
\begin{center}
\hspace{-0.0cm}
\includegraphics[width=0.45\textwidth, trim=0 260 10 10, clip]{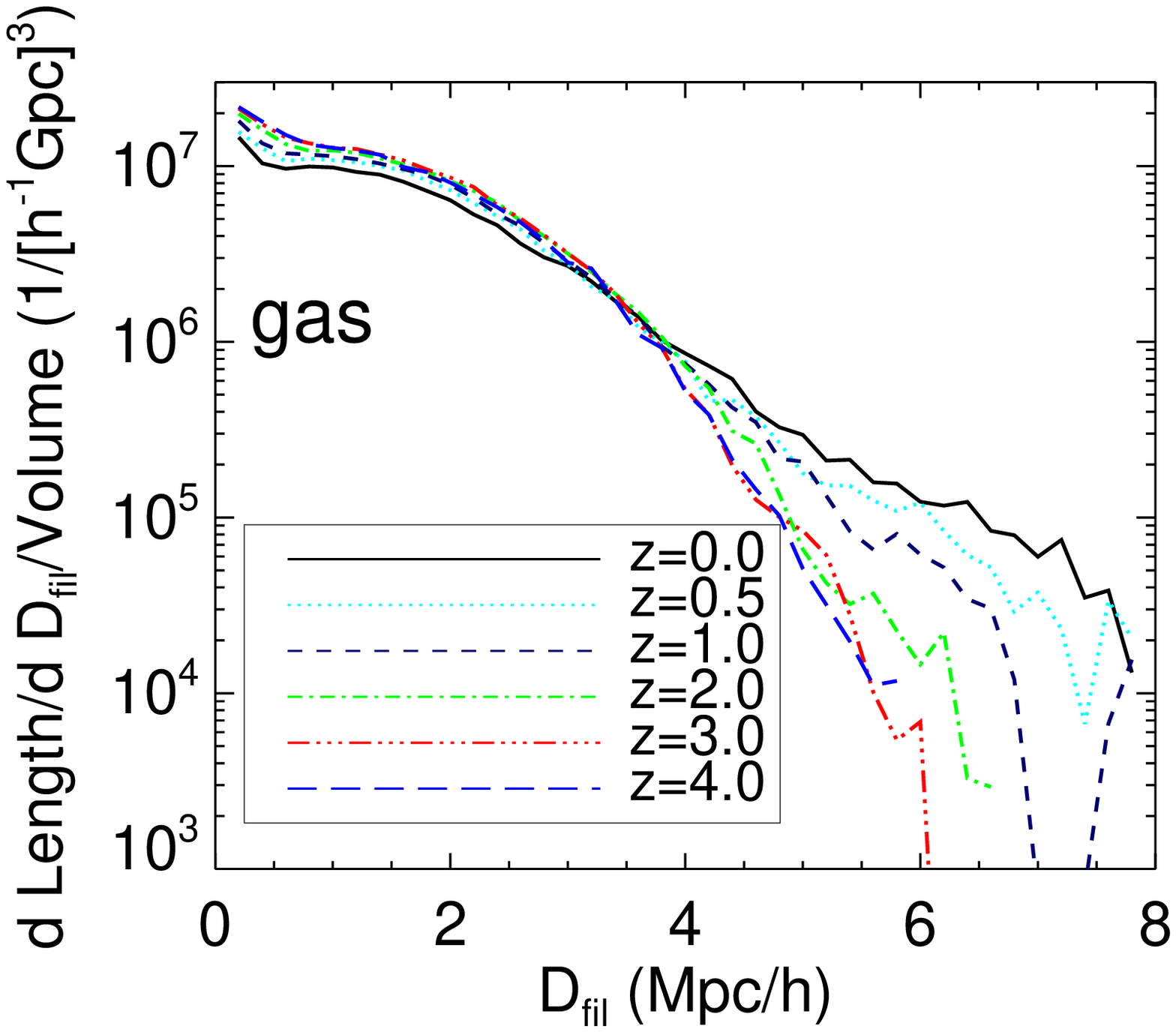}
\includegraphics[width=0.45\textwidth, trim=0 260 10 10, clip]{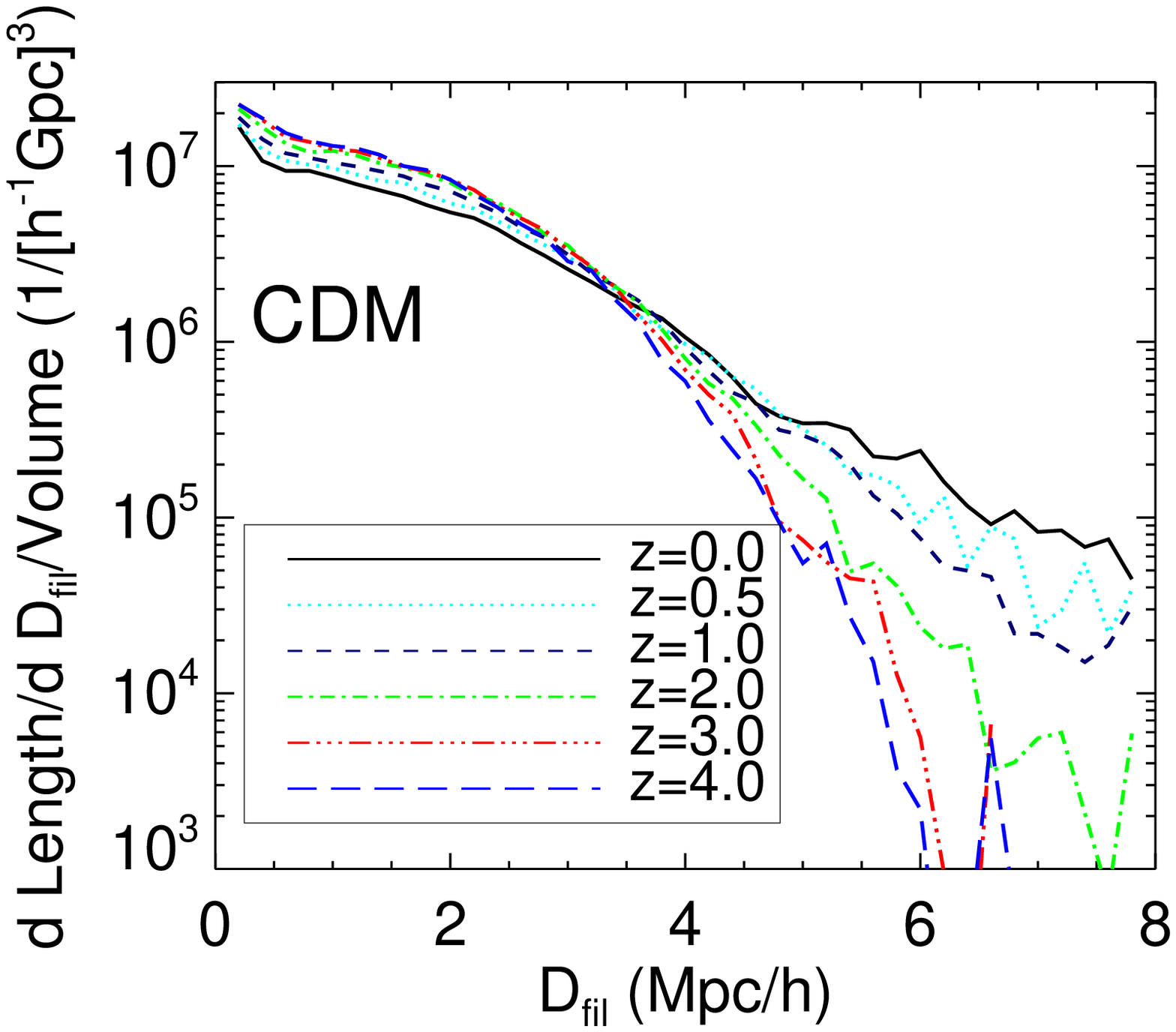}
\includegraphics[width=0.45\textwidth, trim=0 260 10 10, clip]{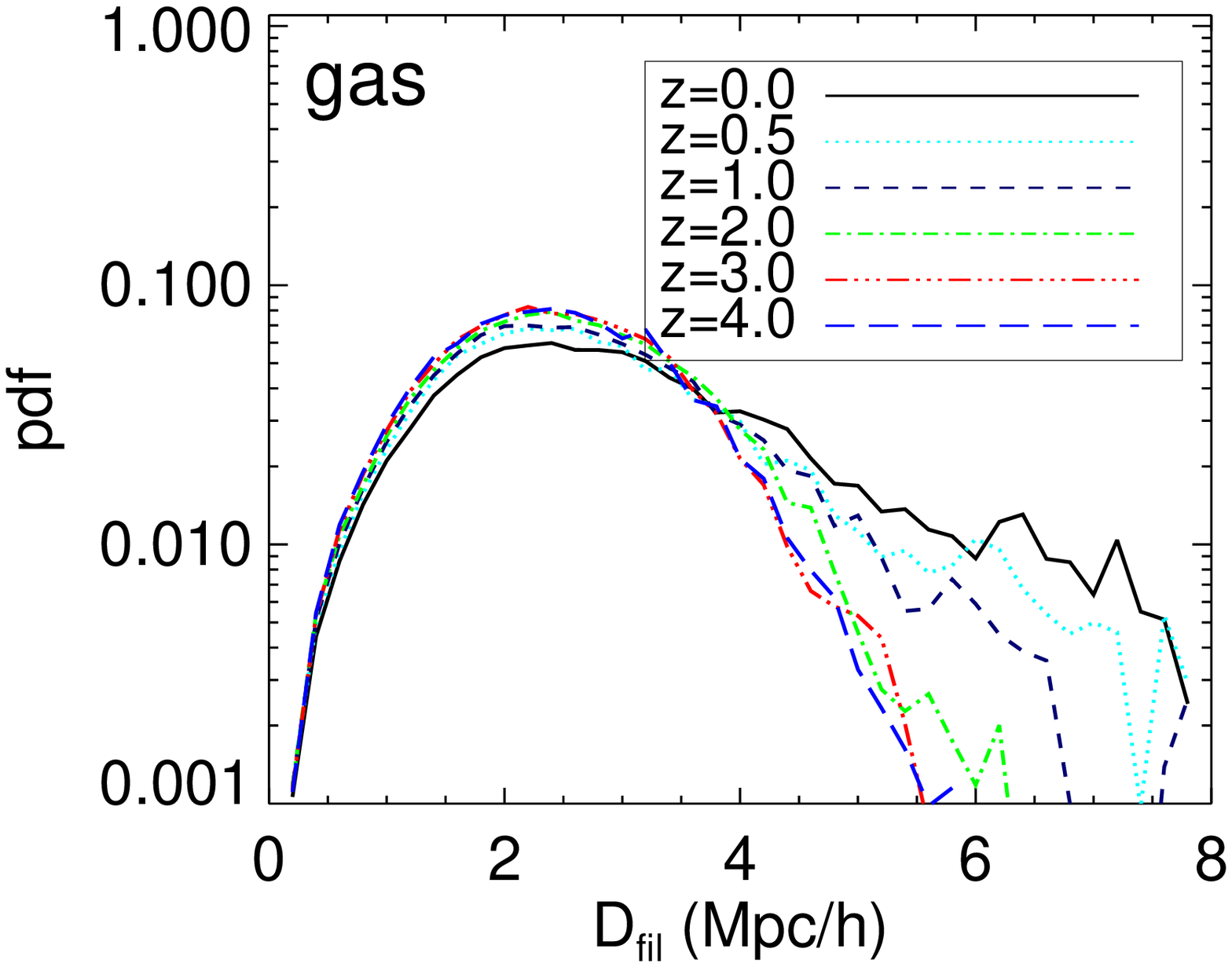}
\includegraphics[width=0.45\textwidth, trim=0 260 10 10, clip]{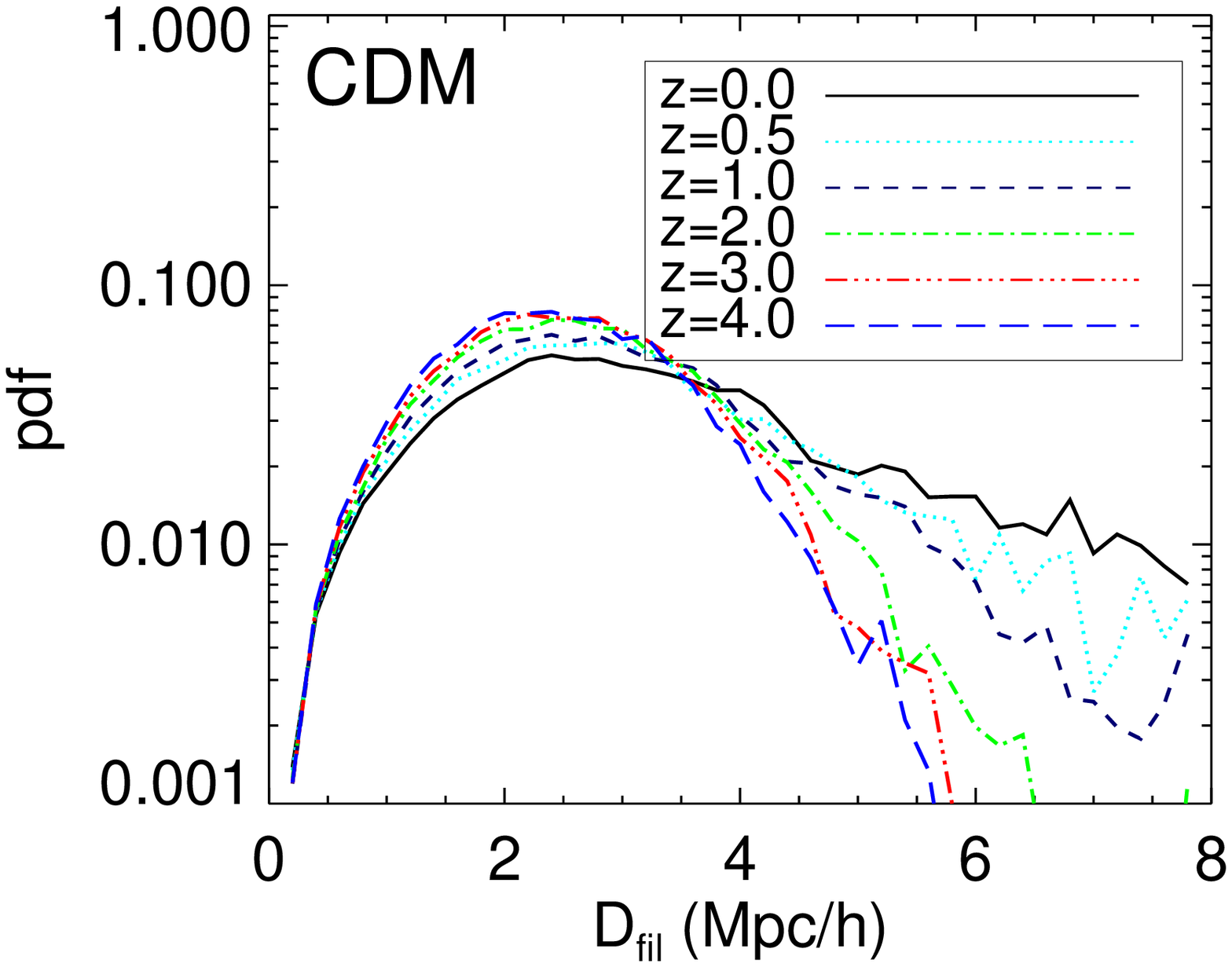}
\includegraphics[width=0.45\textwidth, trim=0 260 10 10, clip]{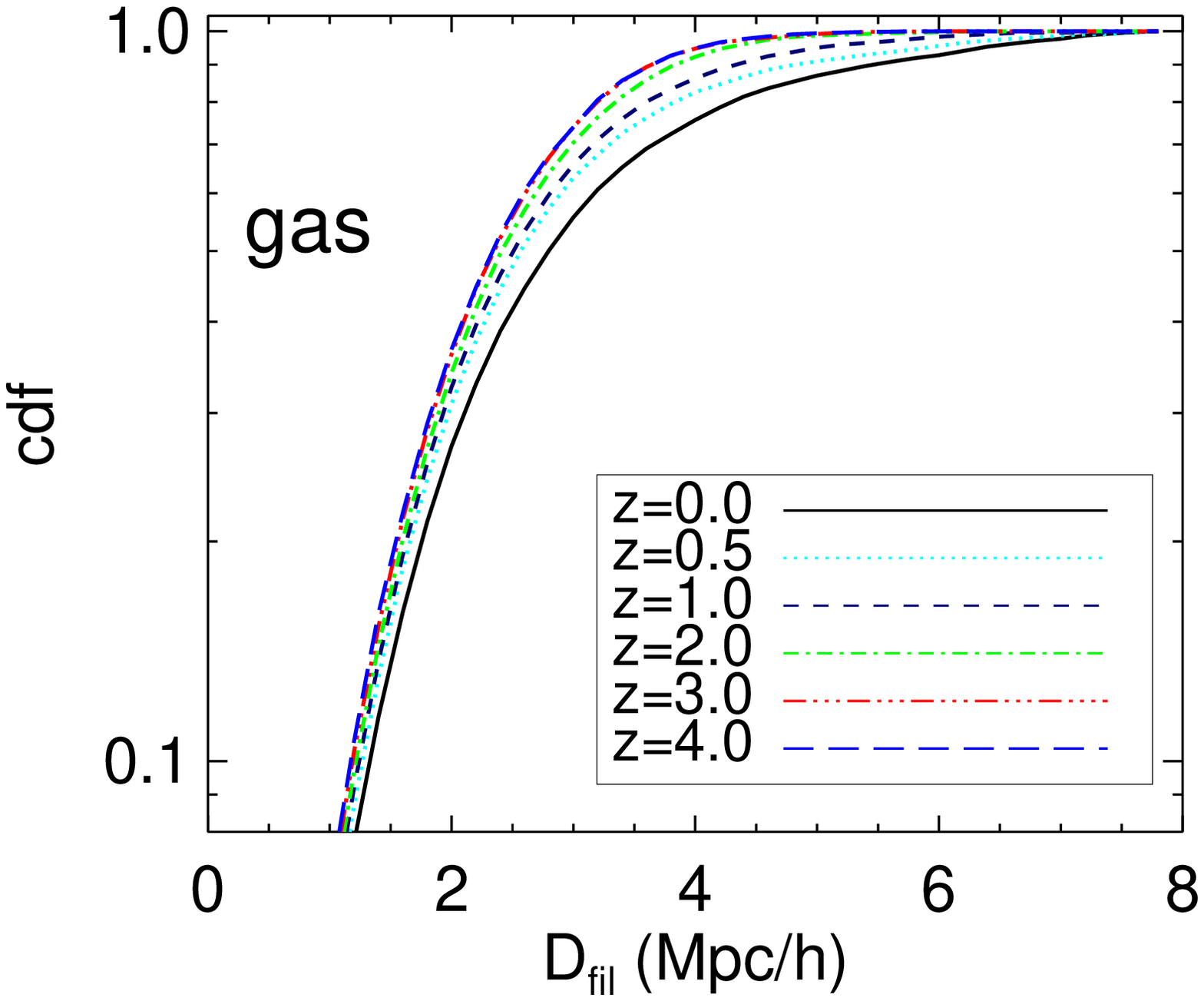}
\includegraphics[width=0.45\textwidth, trim=0 260 10 10, clip]{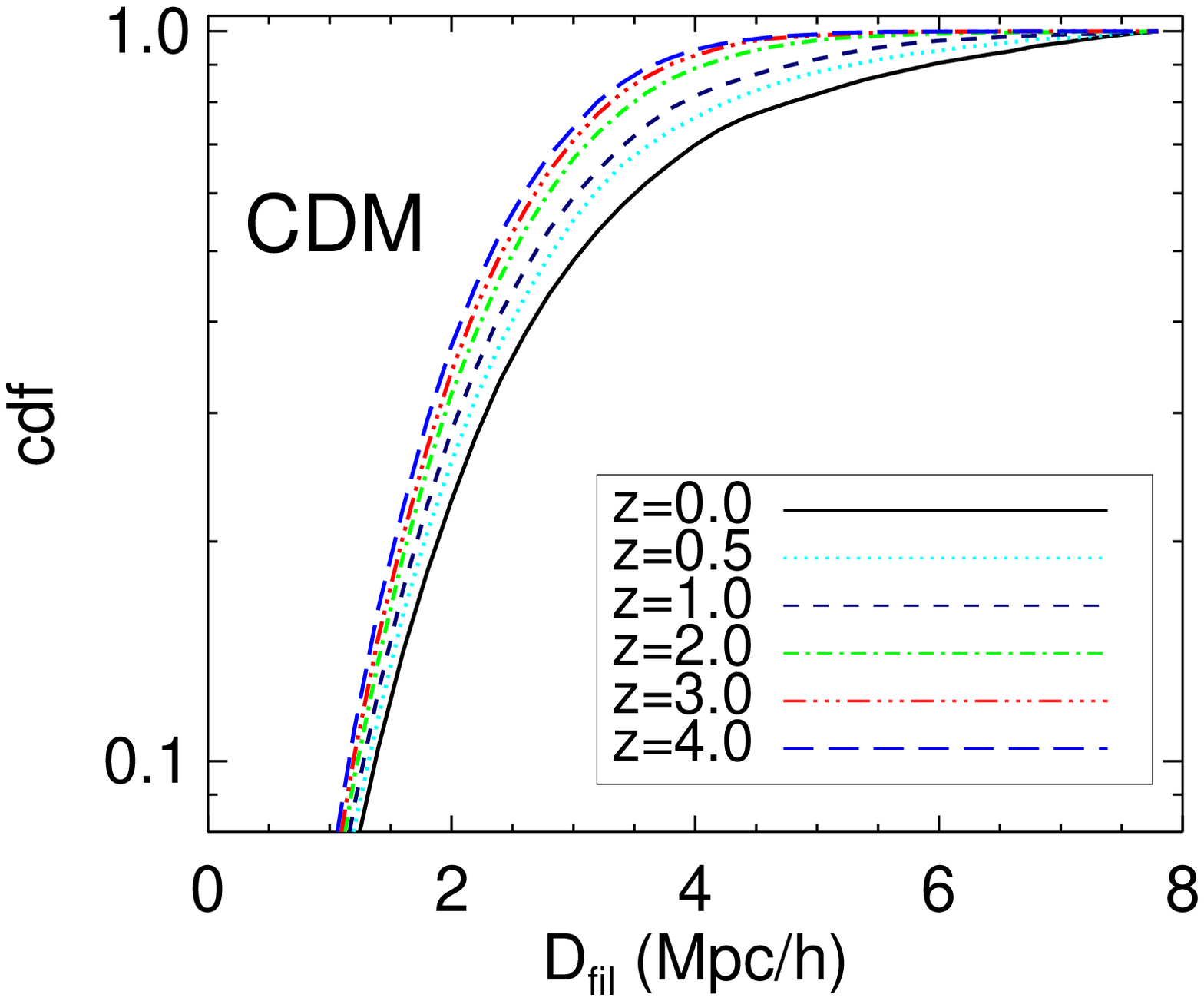}
\caption{Left and right column indicate results of baryonic gas and dark matter filaments respectively. Top: The total length of filaments in each $(Gpc/h)^3$ as a function of the local diameter of filament ($D_{\rm fil}$) at different redshifts. Middle: The probability distribution function of mass as a function of the local diameter. Bottom: The cumulative distribution function of mass residing in filaments as a function of the local diameter.}
\label{fig:fila_freq}
\end{center}
\end{figure*}

\subsection{Width and linear density}
Visual impressions in previous studies have shown that most of the filaments at high redshifts are tenuous ones, and prominent filaments mostly emerging after $z\sim 2$(\citealt{2007PhDT.......196A}; \citealt{2007MNRAS.381...41H}; \citealt{2014MNRAS.441.2923C}; \citealt{2017ApJ...838...21Z}). To estimate this picture quantitatively, we probe the evolution of filament width following the method in \citealt{2014MNRAS.441.2923C}. The top-left and top-right panel in Figure \ref{fig:fila_freq} shows the distribution of the local diameter of baryonic and dark matter filament between $z=4$ and $z=0$ in our samples respectively. The results of dark matter are very similar to baryonic gas. For relatively thinner filaments with local diameter smaller than $\sim 4.0 \rm{Mpc}/h$, the distribution and evolution of filament width is generally consistent with \citealt{2014MNRAS.441.2923C}(see their Figure 38), which is based on samples from the Millennium simulations. However, the number frequency of thicker filaments, with local diameter larger than $4.0 \rm{Mpc}/h$, grows significantly since $z\sim 2$ in our samples, which is in contrast to the trend reported in \citealt{2014MNRAS.441.2923C}. This discrepancy may result from the differences on the methods applied to classify cosmic web. The results in \citealt{2014MNRAS.441.2923C} are based on filaments classified by the NEXUS+ method, which also use the Hessian matrix of density field, but smooth the input density field with a Log-Gaussian filter of multi-radius ranging from $0.4 h^{-1}\, $Mpc to $8.0 h^{-1}\,$ Mpc. In comparison, we smooth the density with a Gaussian filter of fixed radius $0.39 h^{-1}\, $Mpc. In addition, \citealt{2014MNRAS.441.2923C} adopt a threshold value of $\lambda_{th}=0.0$, which is smaller than the value $\lambda_{th}=0.2$ used here.

In our work, the local radius of filaments can be as large as $\sim4$ Mpc. However, few observational work has so far revealed the width of cosmic filaments. Very recently, \cite{2021NatAs.tmp..102W} presents a possible observational evidence for cosmic filament spin based on observed galaxy samples, and claims that the spin signal decreases to zero at a distance to filament spine of $\sim2$ Mpc. Their work seems to suggest that galaxy at a distance $>2$ Mpc to filament spine has been barely impacted by filaments. This may be result from the following reason. The number of filaments with $R_{fil}>2h^{-1}$ Mpc is much less than those with $R_{fil}<2h^{-1}$ Mpc in our sample. Hence the signal induced by filaments with $R_{fil}>2h^{-1}$ Mpc will be relatively weak in stacked filament samples.

As demonstrated in the middle panel of Fig.~\ref{fig:fila_freq}, along with the emerging of prominent filaments, the mass fraction of gas and dark matter residing in thick filaments grows with time. The bottom panel of Fig.~\ref{fig:fila_freq} shows the cumulative distribution function of gas and dark matter mass as a function of the filament local diameter. Filaments with $\rm{D_{fil}}>4.0\, \rm{Mpc}/h$ contain about $7.5\%$ of the IGM residing in filaments at $z=4$. This fraction goes up moderately to about $10.0\%$ at $z=2$, and further increase to about $28\%$ at $z=0$, equivalent to $\sim 13\%$ of all the IGM in the universe. The corresponding mass fraction of dark matter residing in thick filaments is slightly higher than gas at $z\leq 1.0$.  Thin filaments with $\rm{D_{fil}}<2.0\, \rm{Mpc}/h$ comprise $35\%$ of the gas in filaments at $z=0$, and decline to $25\%$ at $z=0$.  

\begin{figure*}[htb]
\begin{center}
\vspace{-3.0cm}
\includegraphics[width=0.45\textwidth, trim=0 10 10 10, clip]{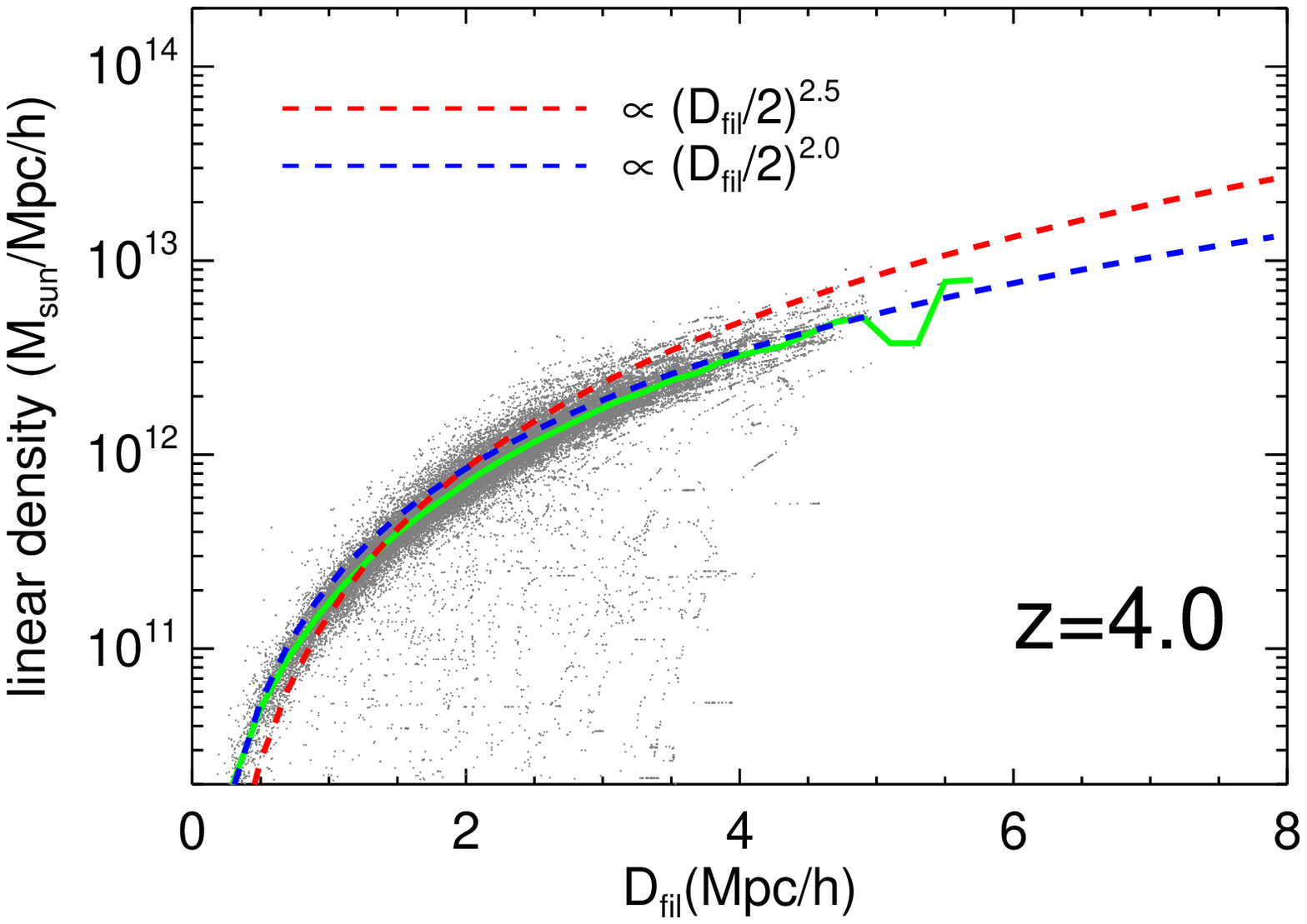}
\includegraphics[width=0.45\textwidth, trim=0 10 10 10, clip]{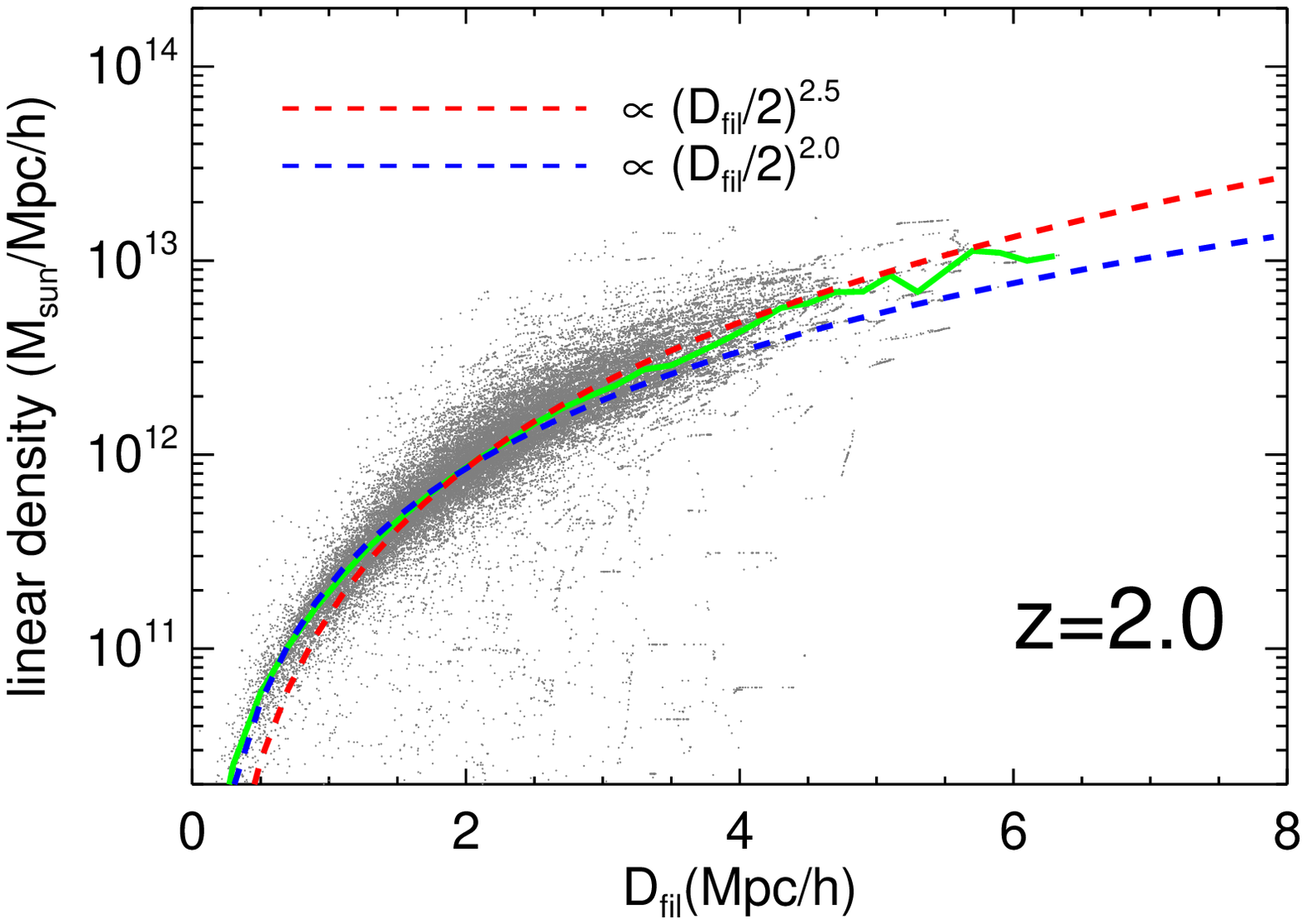}
\includegraphics[width=0.45\textwidth, trim=0 10 10 10, clip]{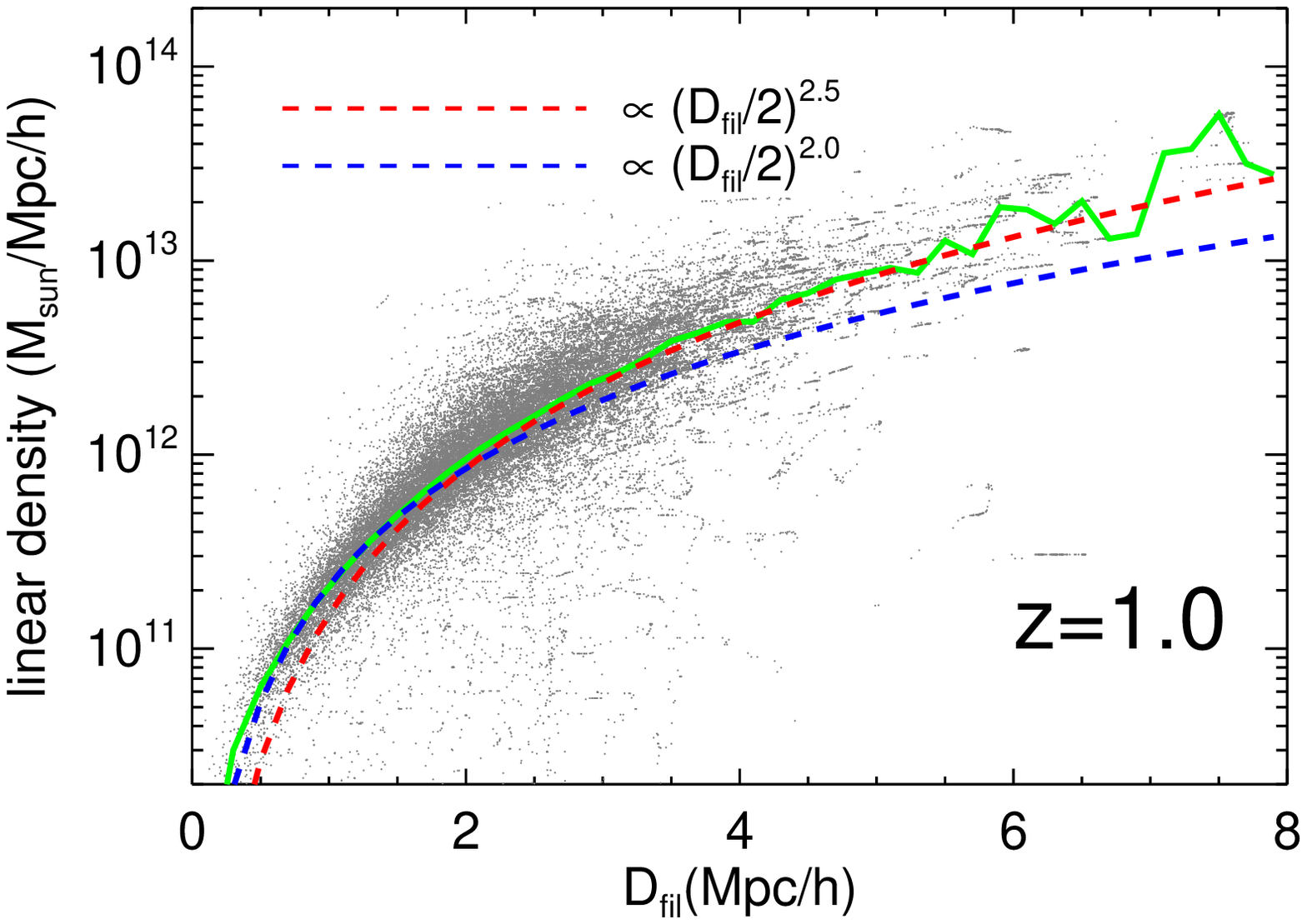}
\includegraphics[width=0.45\textwidth, trim=0 10 10 10, clip]{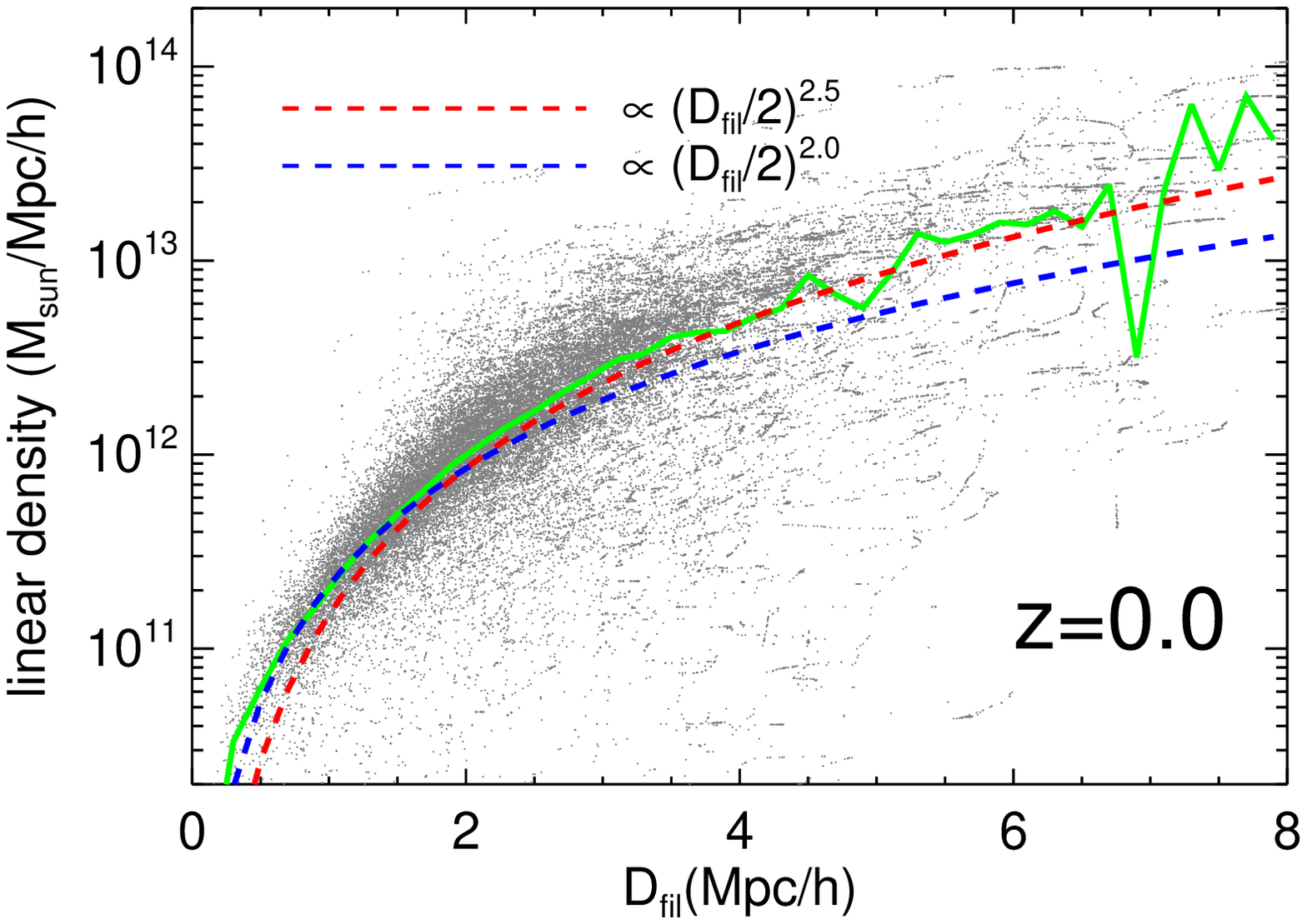}
\caption{The correlation between the local diameter and the linear density of filaments from $z=4.0$ to $z=0.0$. Gray points in each panel are randomly selected 100,000, i.e., about $5\%$ of total, grid cells in filaments. The solid green line indicate the median linear density as a function of $\rm{D}_{fil}$. The red and blue dashed lines show two power law relations.}
\label{fig:fila_linden}
\end{center}
\end{figure*}

\begin{figure*}[htb]
\begin{center}
\hspace{-0.0cm}
\includegraphics[width=0.45\textwidth, trim=0 10 10 10, clip]{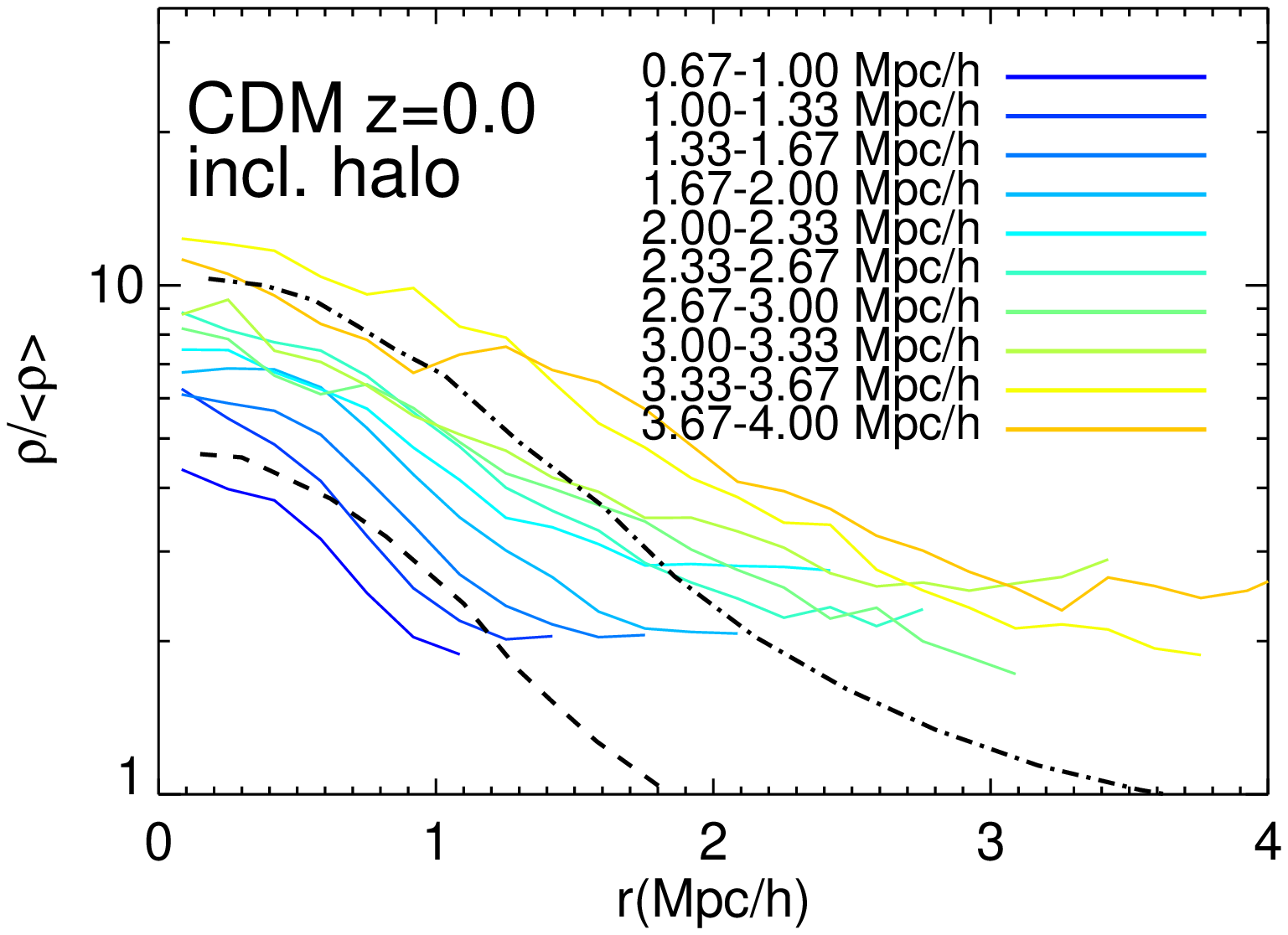}
\includegraphics[width=0.45\textwidth, trim=0 10 10 10, clip]{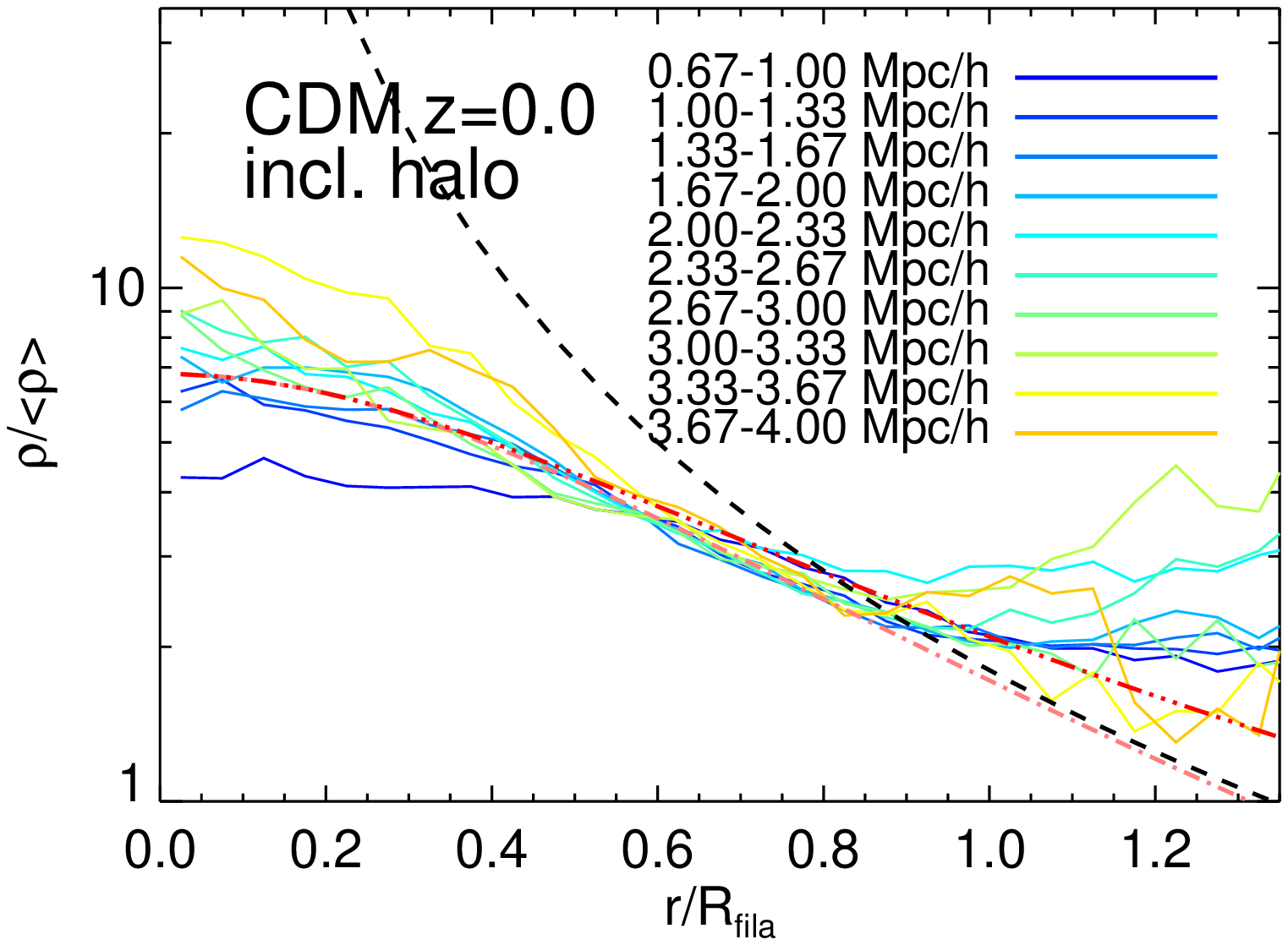}
\includegraphics[width=0.45\textwidth, trim=0 10 10 10, clip]{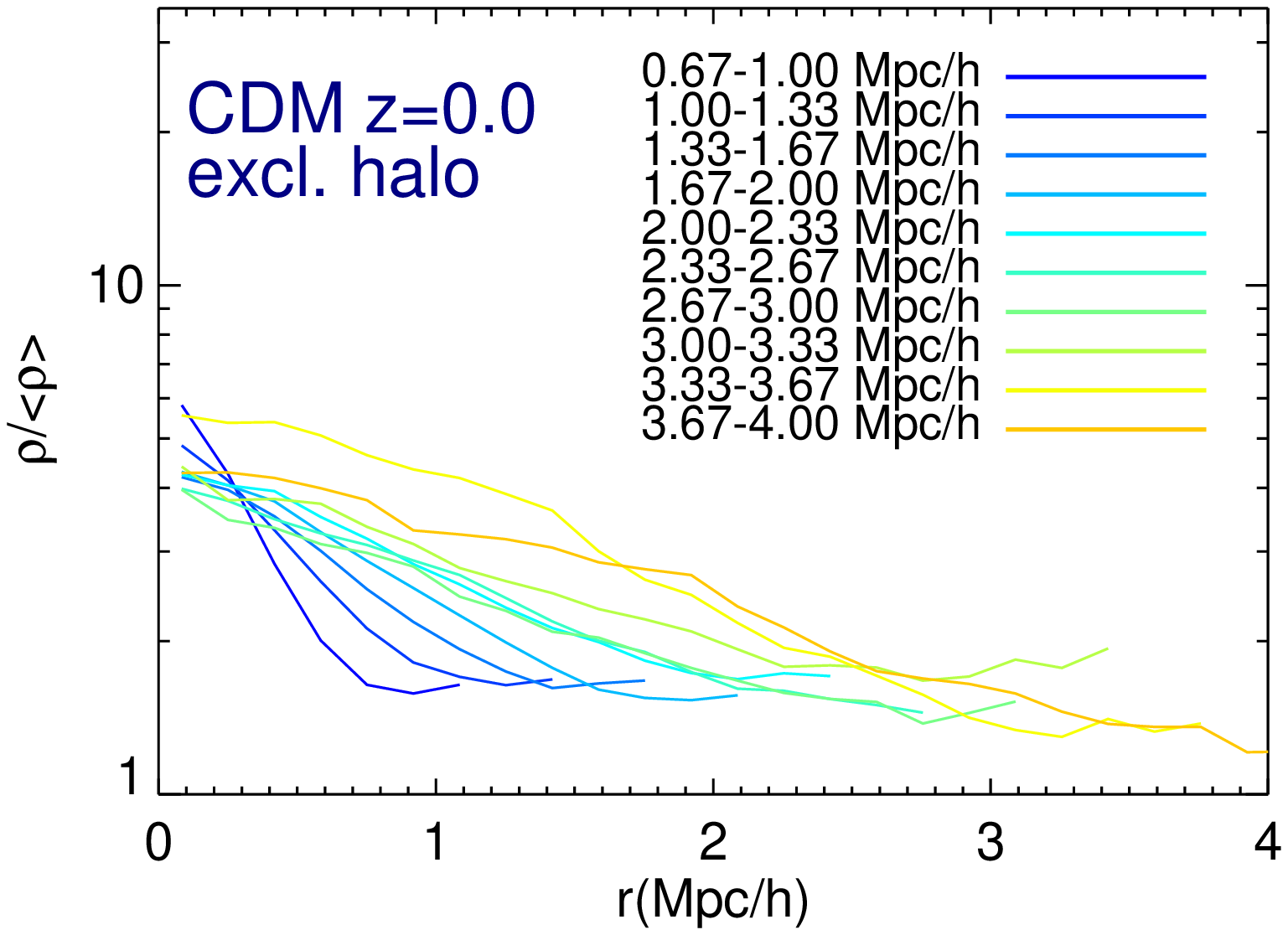}
\includegraphics[width=0.45\textwidth, trim=0 10 10 10, clip]{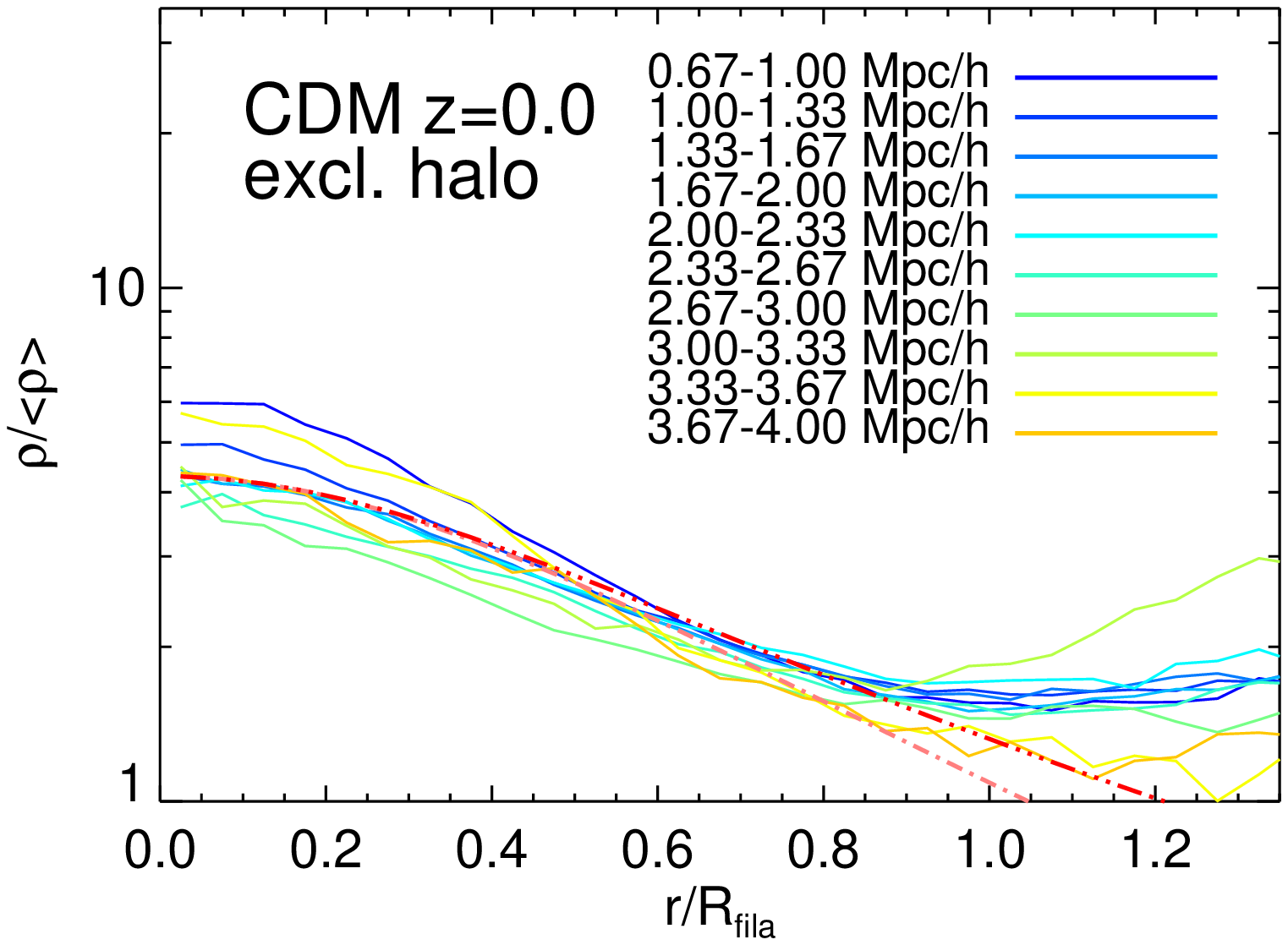}
\caption{Top: The density profile of dark matter as a function of distance(left) and normalised distance(right). to the spine of filament at $z=0$. Different solid lines indicate filaments with particular local radius $R_{fil}$, shown by the legend. Dashed and dotted-dashed black lines in the top-lef panel indicate the profiles of filaments with width $D_{fil}<4.0 \rm{Mpc}/h$ in \cite{2014MNRAS.441.2923C}. Red three-dotted-dashed and pink dotted-dashed lines in the right panel show two single-beta models with $\beta=2/3$ and $\beta=1$ respectively. Black dashed line indicates $\propto 1/(r/R_{fil})^2$. Bottom: The same as top row, but excluding the contribution from collapsed halos. }
\label{fig:filb_denprof}
\end{center}
\end{figure*}

\citealt{2014MNRAS.441.2923C} shows that there is a correlation between the local diameter and the local linear mass density of filaments, $\zeta_{fil}$, although bears with considerable scatter. Fig. ~\ref{fig:fila_linden} presents the scatter diagram between the local linear mass density $\zeta_{fil}$ and local diameter of filaments, identified with the baryon density field from $z=4.0$ to $z=0.0$ in our samples. Note that, the linear mass density measured here is the sum over baryonic and dark matter. Despite there is notable scatter, the local diameter shows a clear trend to correlate with the local linear mass density, in good agreement with \citealt{2014MNRAS.441.2923C}.

The green solid line in Fig. ~\ref{fig:fila_linden} indicates the median linear mass density as a function of the local diameter. We find that this correlation could be approximately fitted by following function as
\begin{equation}
    \zeta_{fil} \approx  \frac{ 2.7\times 10^{11}\rm{M}_{\odot}}{(\rm{Mpc}/h_0)} *\pi*(\frac{R_{fil}}{(\rm{Mpc}/h_0)})^n,
\label{eqn:fila_linden}
\end{equation}
where the power index grows gradually from $n \approx 2$ at $z=4.0$ to $n\approx 2.5$ at $z=0.0$. The increase of power index probably results from the evolution of density profiles in filaments. 

\subsection{Density and thermal profiles}
A well knowledge of the density and temperature profiles of filaments is crucial for the efforts to locate the missing baryon, and for the interpretation of recent observational report of tSZ effect due to filaments. We investigate the density profiles perpendicular to the filament spine in our samples for both baryonic and dark matter, as well as the temperature profiles for baryonic matter. Previous studies have shown that the profiles of individual filaments can vary greatly from one to another(e.g. \citealt{2010MNRAS.408.2163A}, \citealt{2015MNRAS.453.1164G}), and hence profiles are usually calculated by averaging over a group of filaments with certain similar property, such as length( \citealt{2020arXiv201015139G}), or luminosity density(\citealt{2020arXiv201209203T}), or width(\citealt{2014MNRAS.441.2923C}). 

\begin{figure*}[htbp]
\begin{center}
\hspace{-0.0cm}
\includegraphics[width=0.45\textwidth, trim=0 10 10 10, clip]{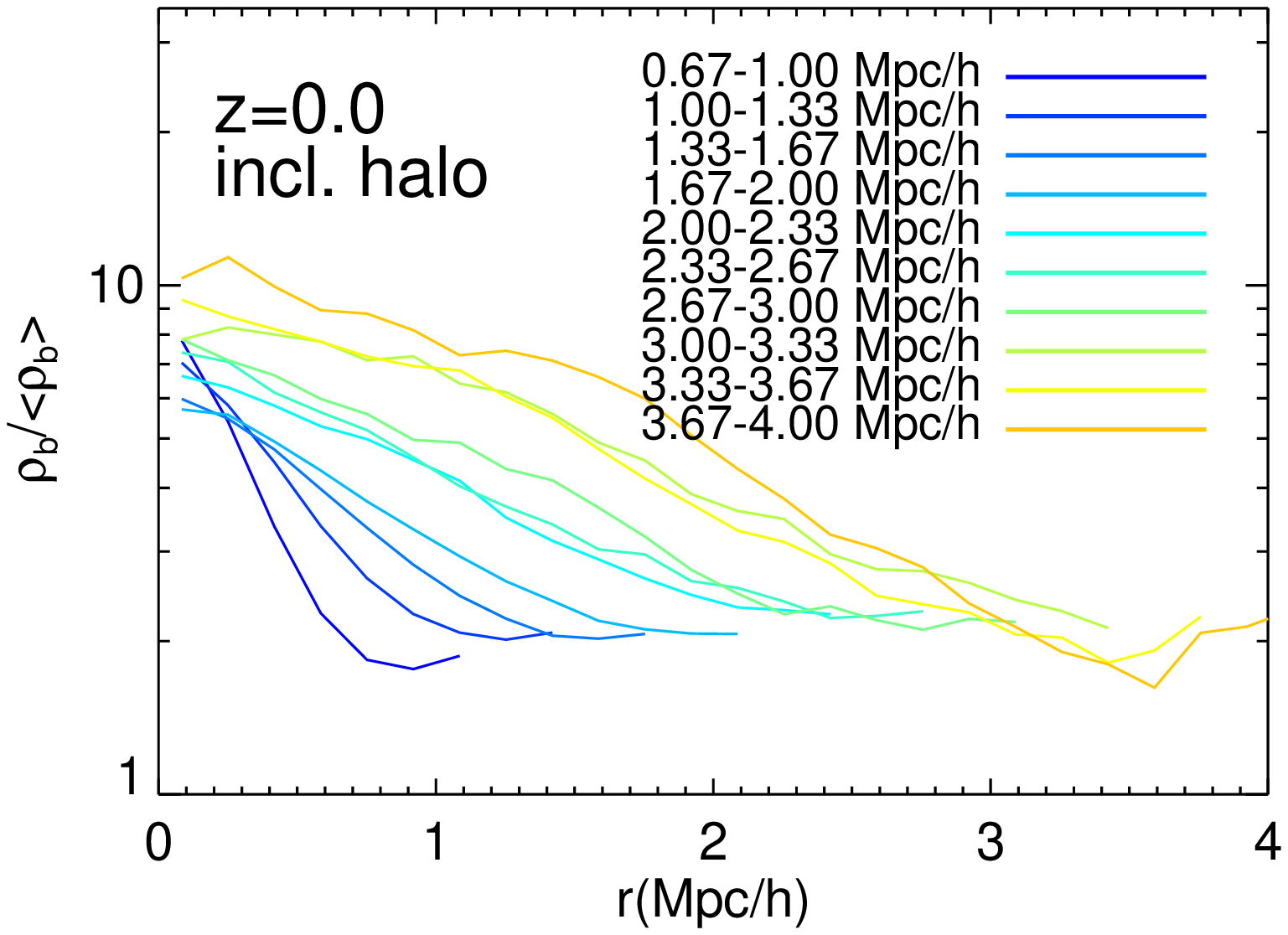}
\includegraphics[width=0.45\textwidth, trim=0 10 10 10, clip]{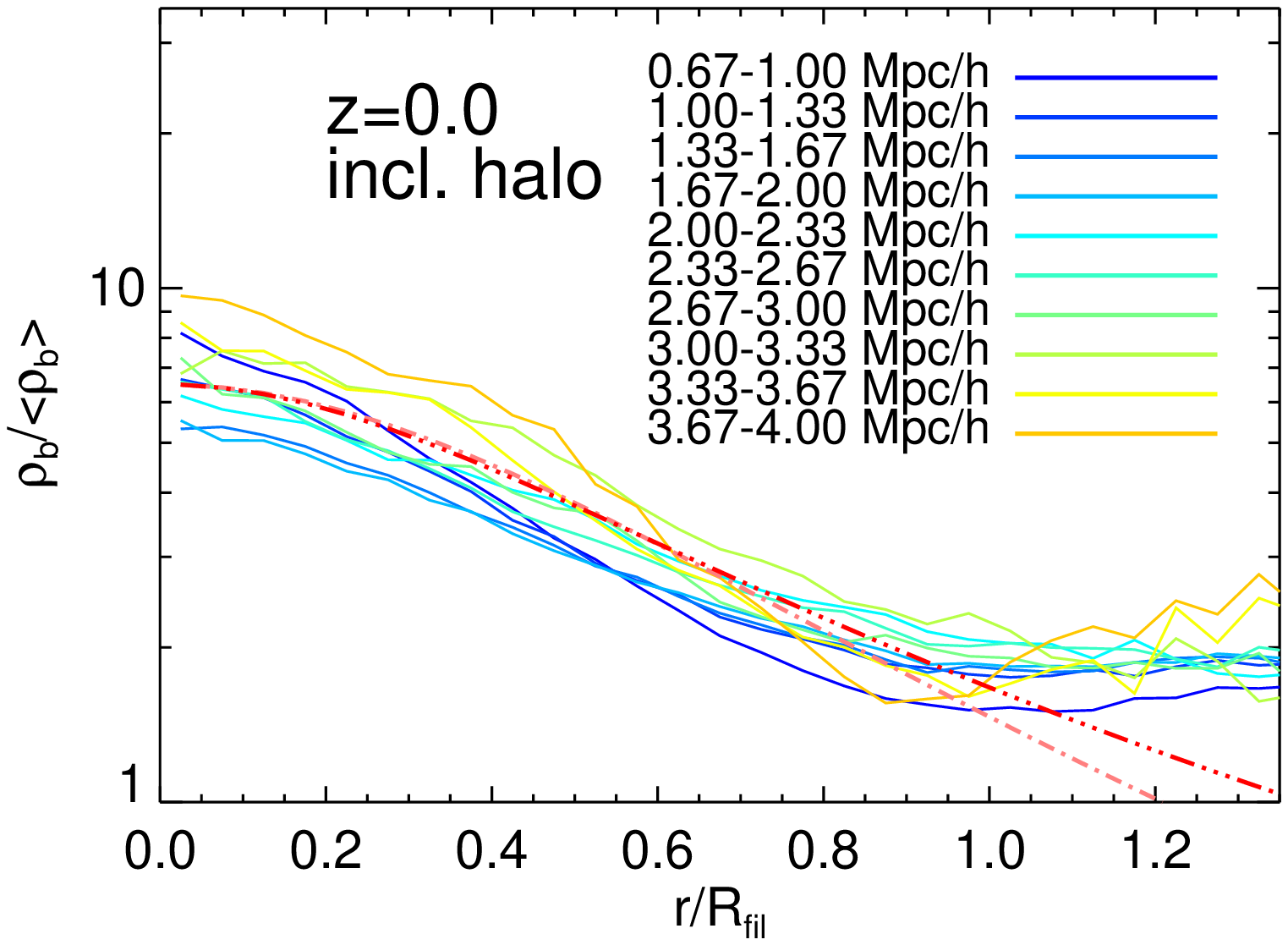}
\includegraphics[width=0.45\textwidth, trim=0 10 10 10, clip]{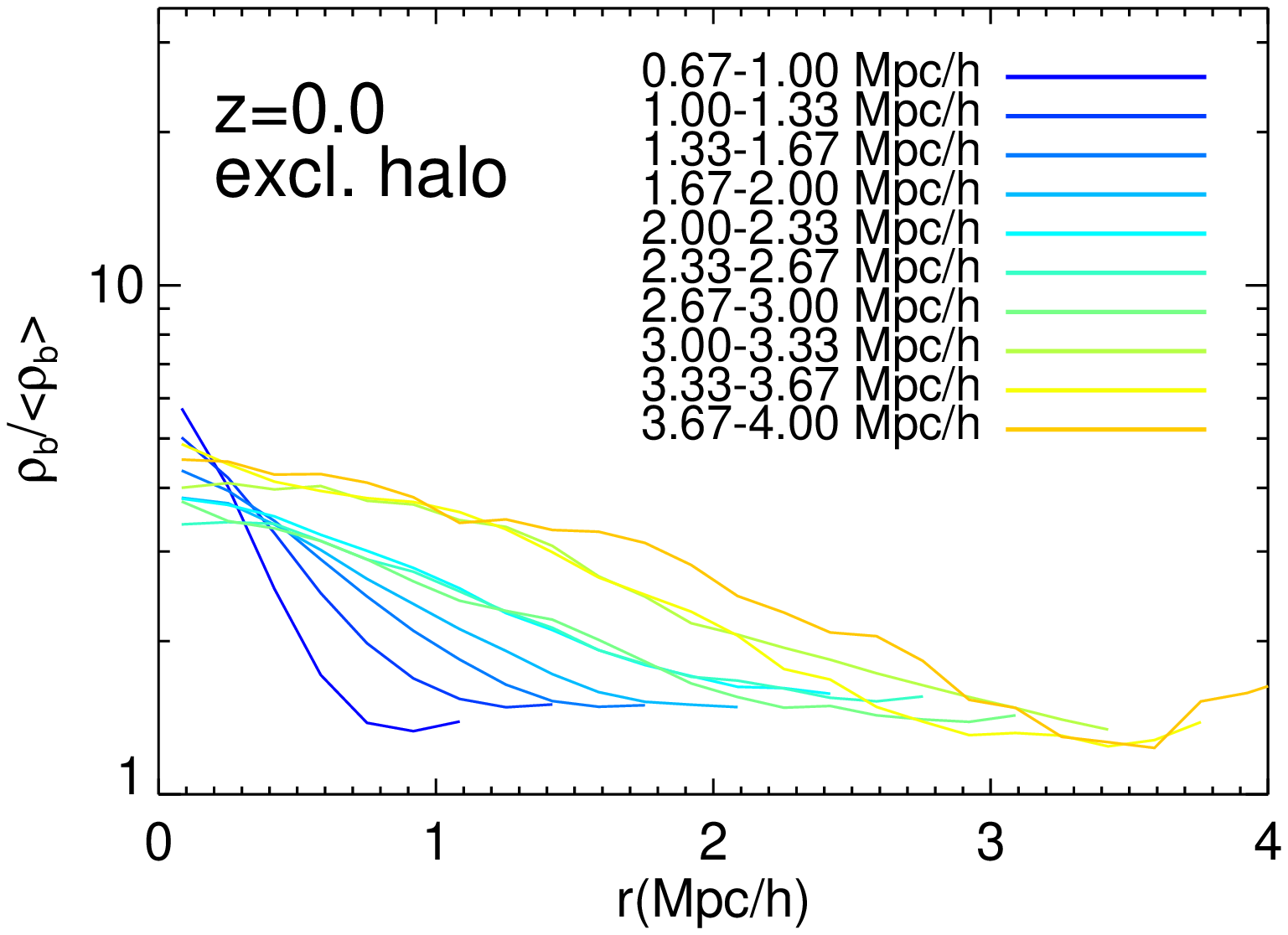}
\includegraphics[width=0.45\textwidth, trim=0 10 10 10, clip]{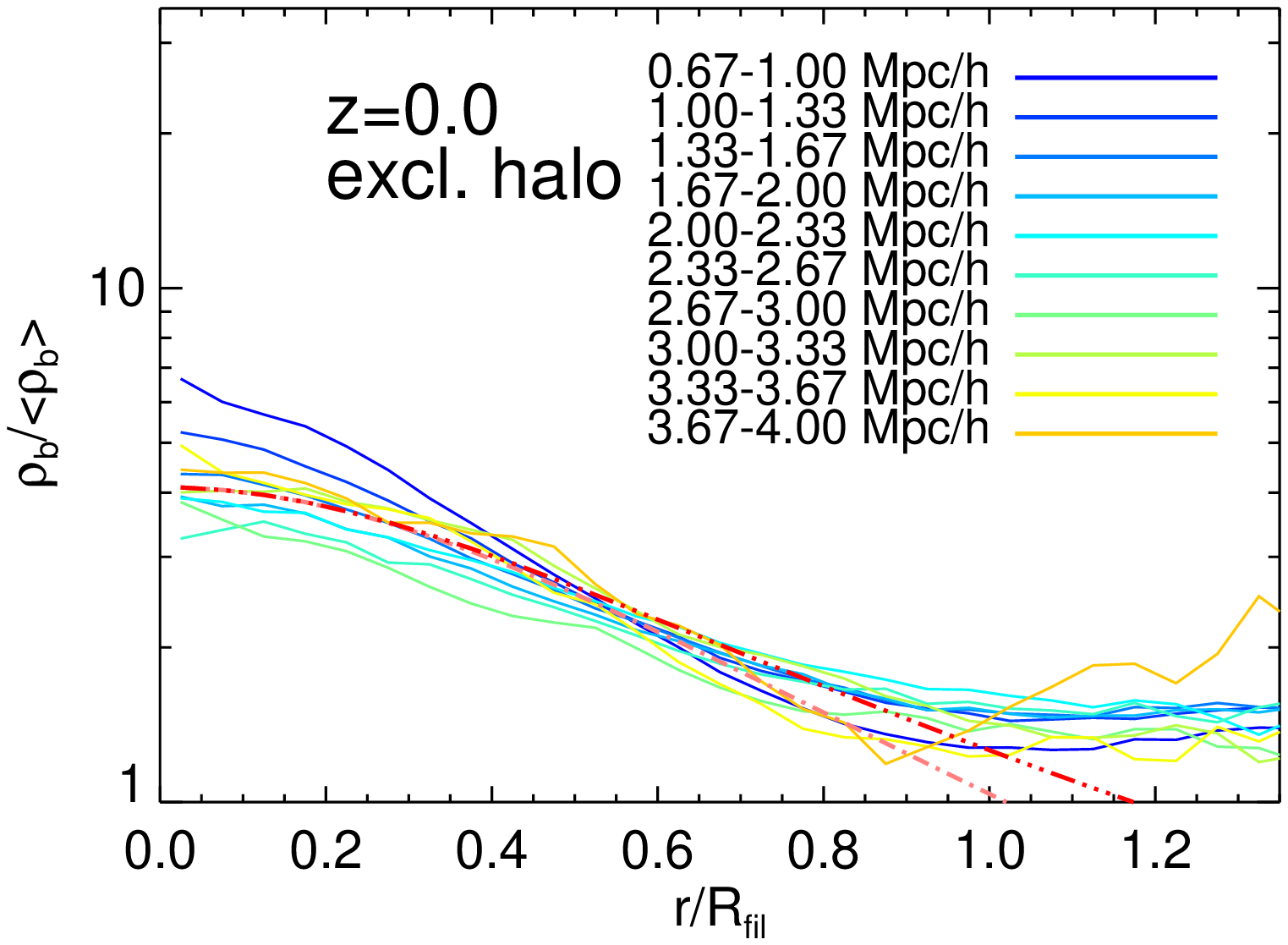}
\caption{Top: The density profile of baryon gas as a function of distance(left) and normalised distance(right) to the spine of filament.Different lines indicate filaments in different width bin. Bottom: The same as top row, but excluding the contribution from collapsed halos.}
\label{fig:fila_denprof}
\end{center}
\end{figure*}

Here, we measure the average profiles of filament segments with similar local width/diameter. More specifically, all the grid cells in filaments are firstly separated to subgroups according to their local diameters. We use an interval of $0.66 \rm{Mpc}/h$ on local diameter as the bin size of subgroup. This is equivalent to a bin size of $0.33 \rm{Mpc}/h$ on local radius as $R_{fil}=D_{fil}/2$. Then we calculate the average density and temperature of cells as a function of the radial distance to the spine $r$ in each subgroup. Both volume weighted and mass weighted temperature have been calculated. The former and latter are defined as $(\sum_{i=1}^{n} T_i)/n$, and $(\sum_{i=1}^{n} T_i \rho_i)/(\sum_{i=1}^{n} \rho_i)$, where $T_i$ and $\rho_i$ are the temperature and density of ith gas cells respectively, and n is the number of gas cells in a particular subgroup. Since there is a small number of filaments with $R_{fil}>4.0 \, \rm{Mpc}/h$ in our sample, our study focus on filaments with local radius $R_{fil}<4.0 \, \rm{Mpc}/h$. Since we only include grid cells belonging to filaments in the calculation, the profiles are truncated at a radius $r \approx R_{fil}$. 

The top-left panel in Fig.~\ref{fig:filb_denprof} shows the density profiles of dark matter in filaments at $z=0$. The overall shape and overdensity of the profiles in our samples are similar to those filaments with a width $D_{fil}<4.0 \rm{Mpc}/h$ in \cite{2014MNRAS.441.2923C}, which shown as the black dashed and dotted-dashed lines in the top-lef panel.  The density profiles can be roughly divided to an inner core regime at $r<0.5R_{fil}$ and an outer envelop regime for filaments with different widths. The peak of the density in the inner regions increases with the local radius of the filament. As the distance to the spine increase, the matter density first declines mildly in the inner core regime and then drops more rapidly to $r=0.8-1.0 R_{fil}$, and finally flatten out at larger radius with some fluctuations. The plateau at $r\sim 1.0 R_{fil}$ is mainly due to grids only in filaments are averaged to obtain the profiles, while grids with lower density, belonging to walls and voids, are not included. Furthermore, we calculate the density profiles as a function of the normalised radial distance to the spine, i.e. $r_{nml}=r/R_{fil}$, shown by the top-right panel in Fig.~\ref{fig:filb_denprof}. For filaments with $R_{fil}$ between $1.0$ and $4.0\, \rm{Mpc}/h$, the density profiles present feature of self-similarity, and can be approximately fitted by an isothermal single-$\beta$ model as
\begin{equation}
\rho_{dm}(r/R_{fil})=\rho_{dm,0} \times (1+(\frac{r}{r_c})^2)^{-\frac{3}{2} \beta_{dm}}
\label{eqn:den_cdm}
\end{equation}
with $\rho_{dm,0}=6.8*(\Omega_m-\Omega_b)*\rho_{crit,0}$, $r_c=0.80R_{fil}$ and $\beta_{dm}=2/3$. The profiles within $r<0.8R_{fil}$ can also fitted with the same $\rho_{dm,0}$ and $r_c$ but a larger $\beta=1.0$, displayed as the pink dotted-dashed line. Yet, the profiles of filaments thinner than $R_{fil}=1.0 \rm{Mpc}/h$ or thicker than $R_{fil}=3.33 \rm{Mpc}/h$ deviate evidently from the single-beta model in the inner core regime. The former and latter have profiles are more shallow and steeper than Eqn.~\ref{eqn:den_cdm} respectively. 

Note that, collapsed halos could make considerable contributions to the matter density in the inner regime of filament. To exclude the influence of halos, we discard those grid cells located within spheres centred on the mass center of halos massive than $6.2 \times 10^8 \rm{M}_{\odot}$ and with radius of $1.2$ times of halos' virial radius. As a result, the measured overdensity in filaments drops moderately in the inner core regime and slightly in the outer envelop regime. It is not surprising. The density profiles of filaments excluding contributions from halos are illustrated in the bottom row of Fig.~\ref{fig:filb_denprof}. Once the contributions from halos are excluded, the density profiles can be better fitted by the isothermal single-$\beta$ model as Eqn.~\ref{eqn:den_cdm} with parameters $\rho_{dm,0}=4.3*(\Omega_m-\Omega_b)*\rho_{crit,0}$, $r_c=0.8R_{fil}$ and $\beta_{dm}=2/3$.

\begin{figure*}[htbp]
\begin{center}
\hspace{-0.0cm}
\includegraphics[width=0.45\textwidth, trim=0 10 10 10, clip]{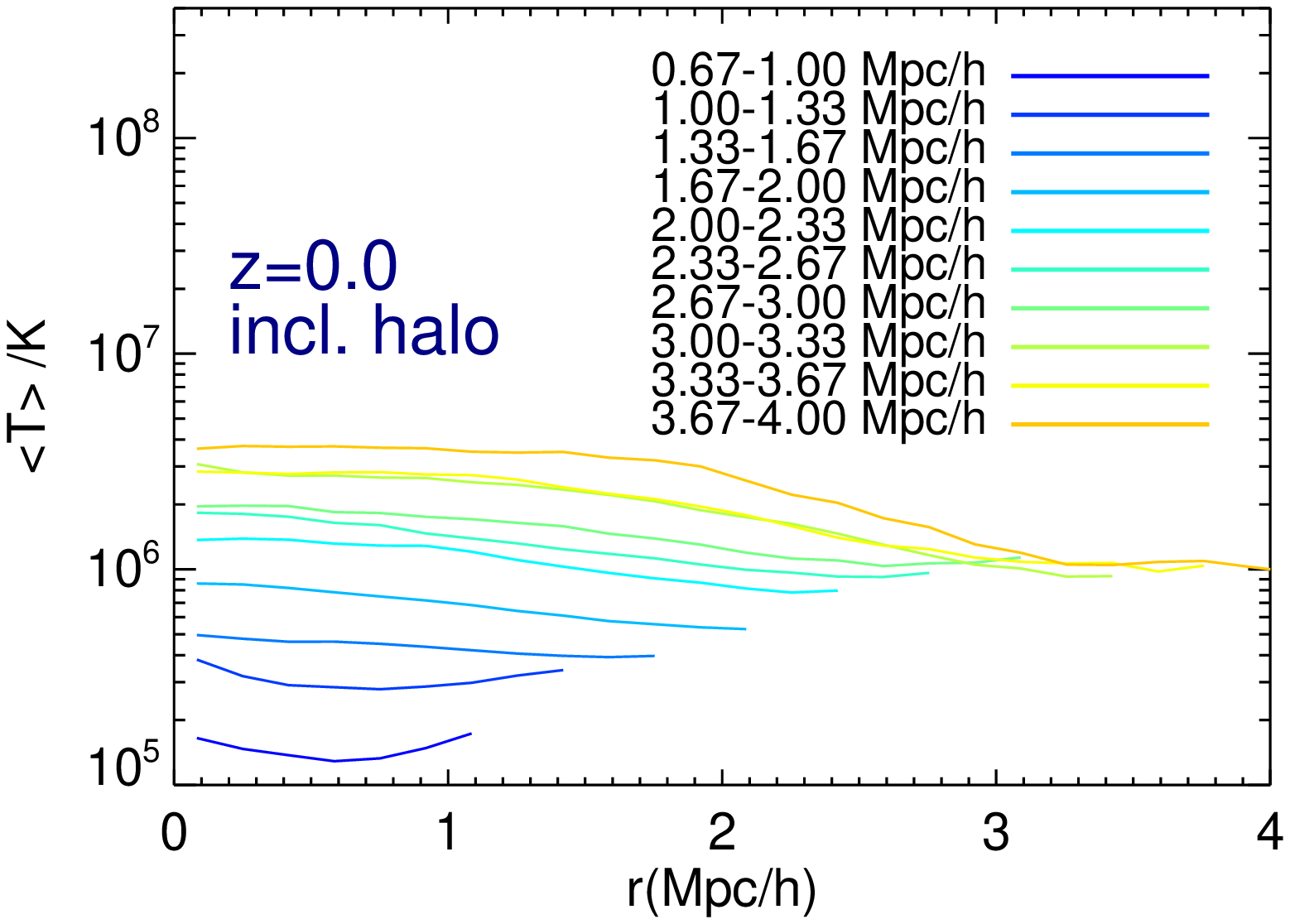}
\includegraphics[width=0.45\textwidth, trim=0 10 10 10, clip]{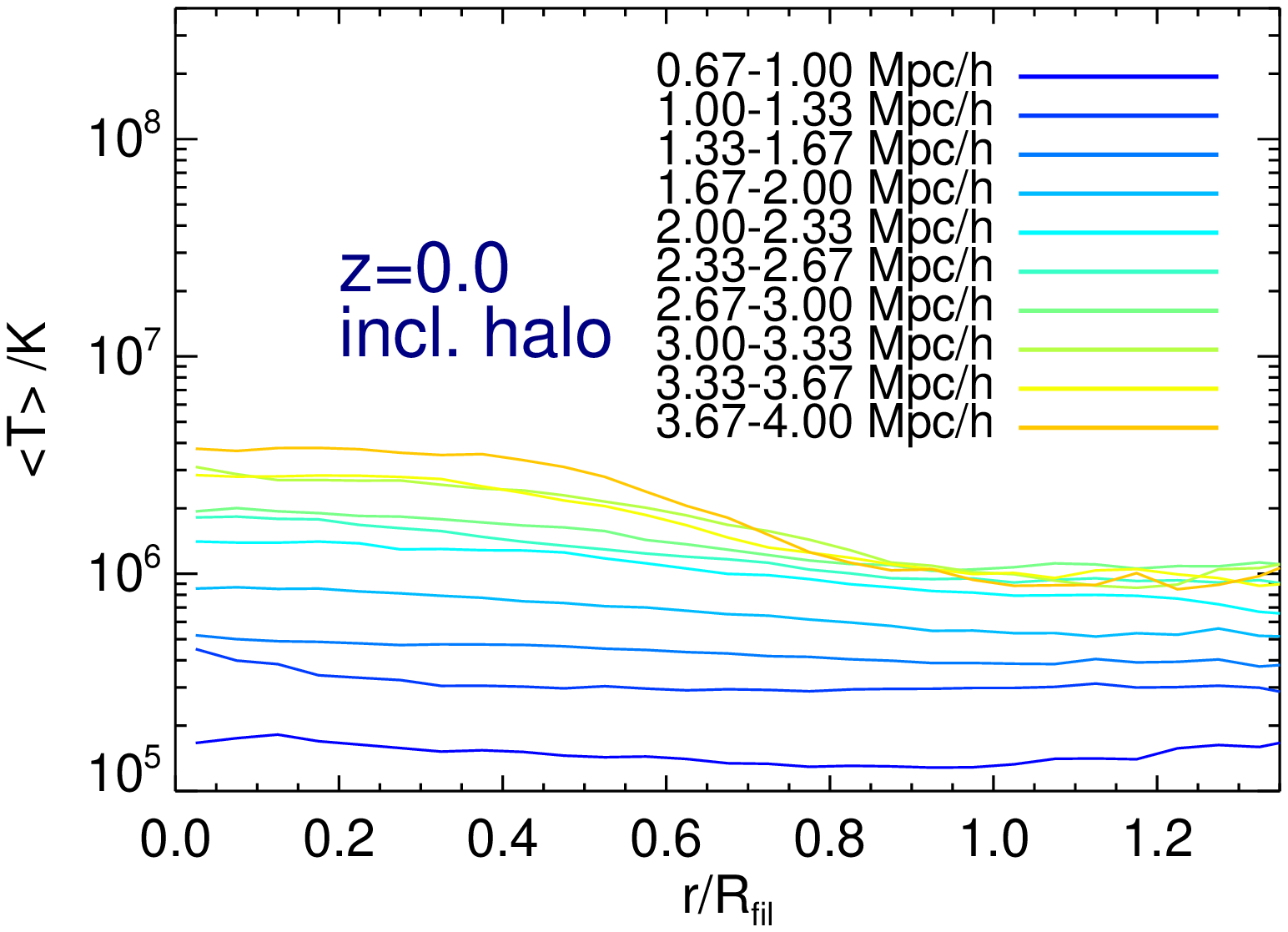}
\includegraphics[width=0.45\textwidth, trim=0 10 10 10, clip]{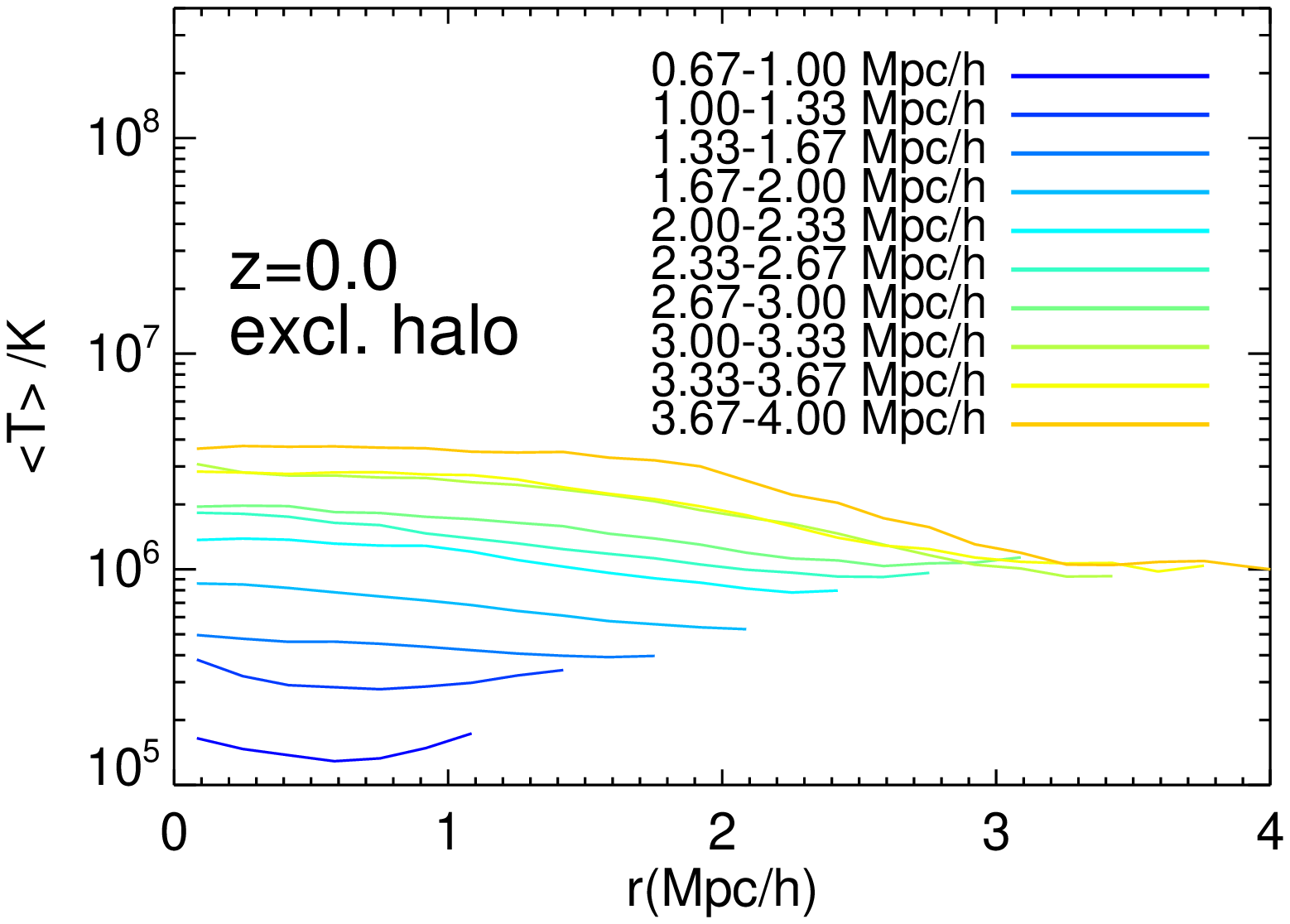}
\includegraphics[width=0.45\textwidth, trim=0 10 10 10, clip]{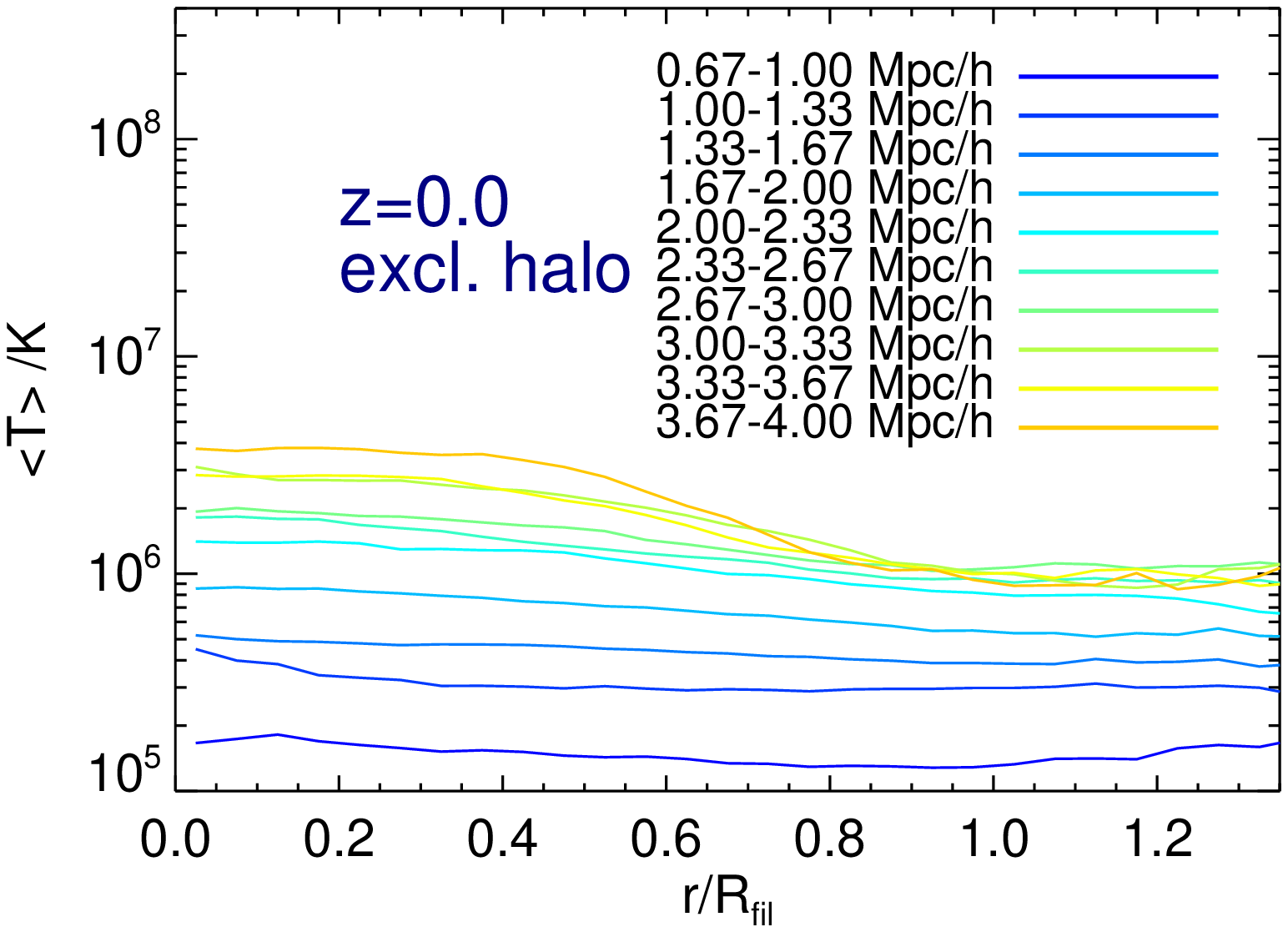}
\includegraphics[width=0.45\textwidth, trim=0 10 10 10, clip]{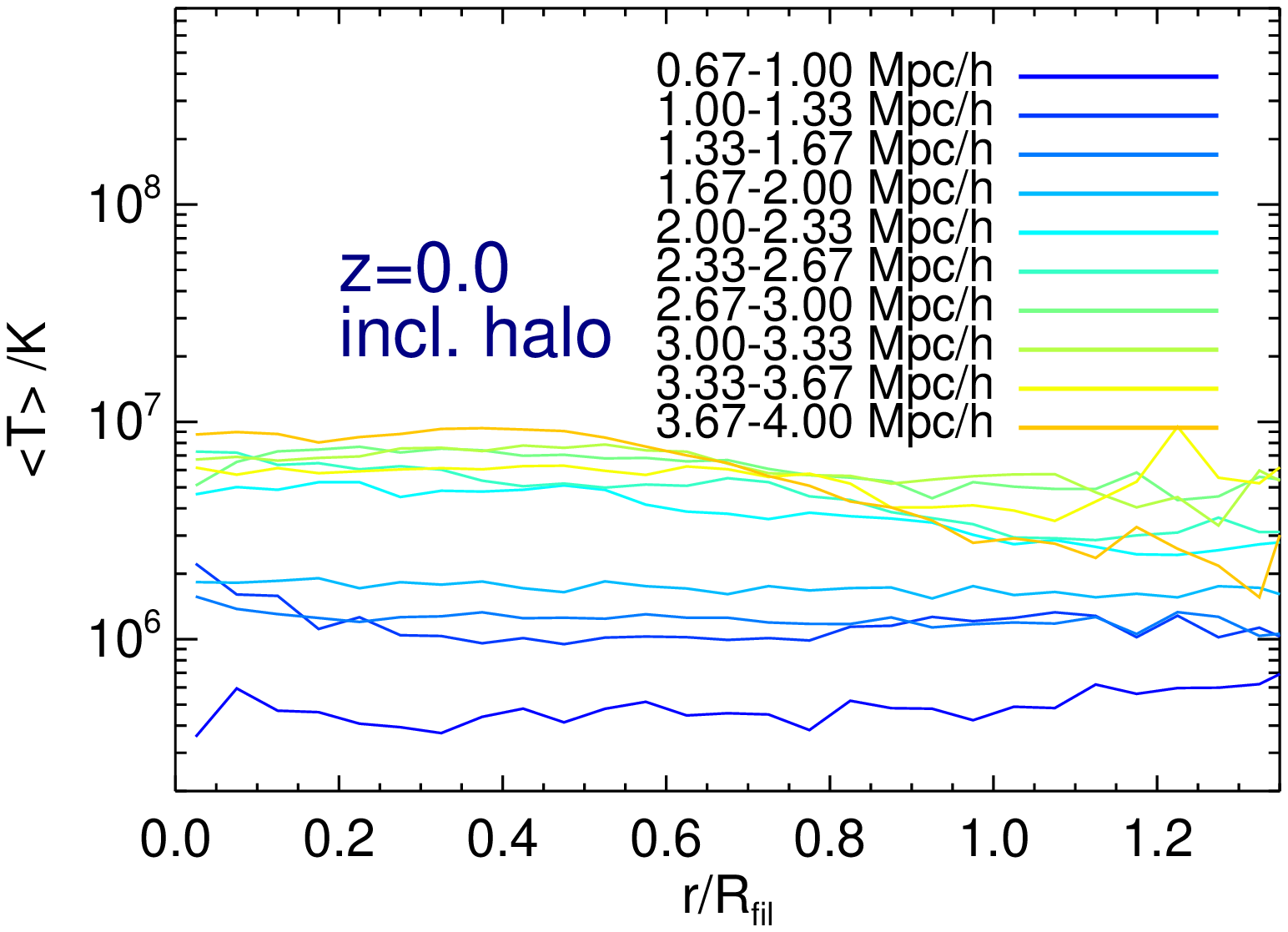}
\includegraphics[width=0.45\textwidth, trim=0 10 10 10, clip]{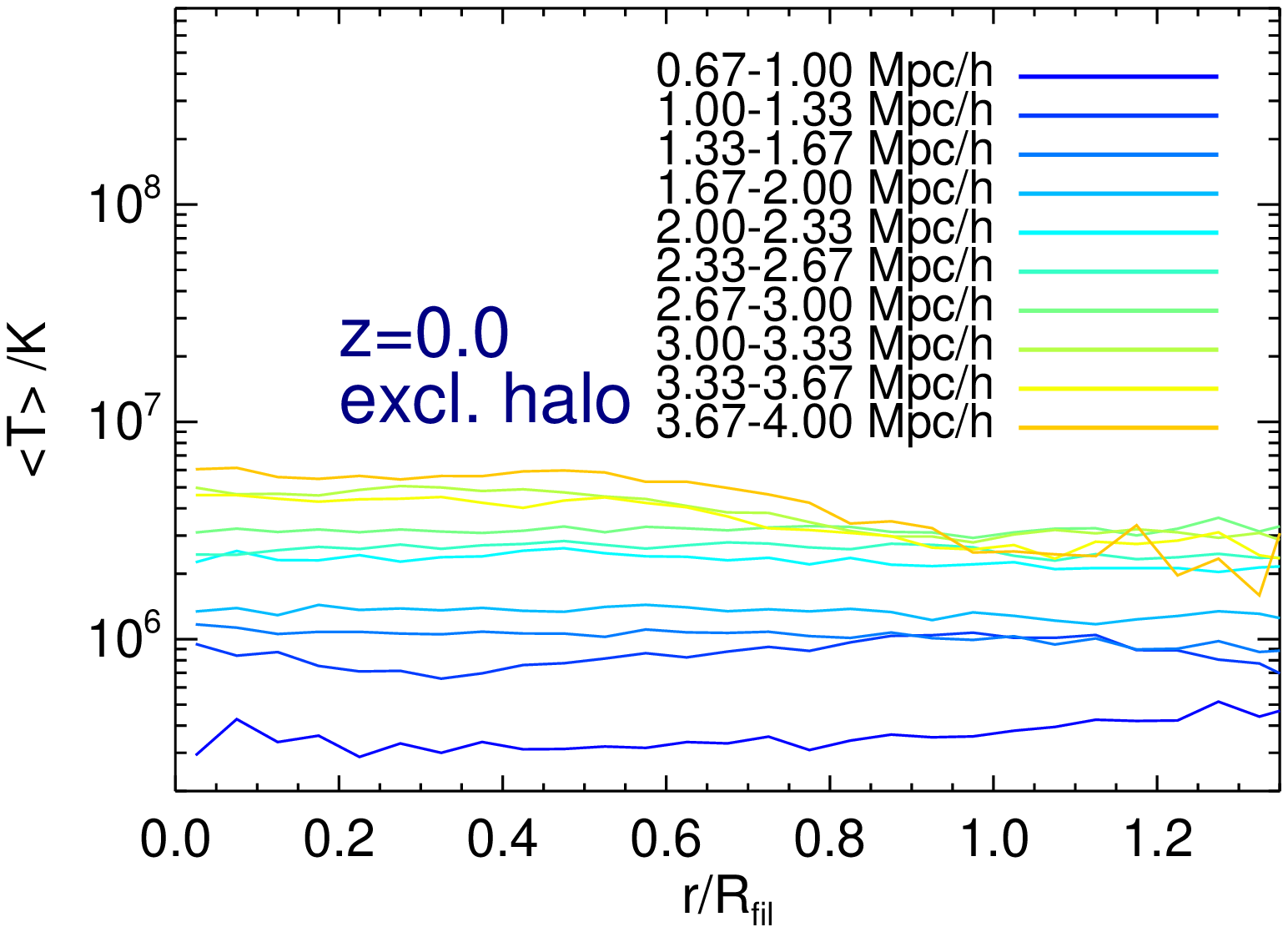}
\caption{Top: The volume weighted temperature profile of baryon gas as a function of distance(left) and normalised distance to the spine of filament. Middle: The same as the top row, but excludes the contribution from collapsed halos. Bottom: The mass weighted temperature profile of baryon gas in filament including(left)/excluding(right) the contribution from halos.}
\label{fig:fila_temprof}
\end{center}
\end{figure*}

Fig.~\ref{fig:fila_denprof} shows the density profiles of baryonic gas in filaments at $z=0$, which are generally similar to those of dark matter and can also approximately fitted by an isothermal single-$\beta$ model as
\begin{equation}
\rho_{gas}(r/R_{fil})=\rho_{gas,0} \times (1+(\frac{r}{r_c})^2)^{-\frac{3}{2} \beta_{gas}}
\label{eqn:den_gas}
\end{equation}
with $\rho_{gas,0}$ equals to 6.5 and 4.1 times of the cosmic mean baryon density if the contributions from halos are included and excluded respectively, being slightly lower than the corresponding overdensity of dark matter. Meanwhile $r_c=0.8R_{fil}$ and $\beta_{gas}=2/3$, which are the same as dark matter.

Fig.~\ref{fig:fila_temprof} illustrates the volume weighted temperature profiles of baryonic gas in filaments at $z=0$. Baryonic gases residing in thick filament are generally hotter than those in thin ones. Filaments with $R_{\rm fil}<2.0\, \rm{Mpc}/h$ are cooler than $10^6 K$, and vice versa. When moving from the boundary region inward to the central region along the direction perpendicular to the spine, the gas temperature shows different behavior in filaments with different width. In filaments thinner than $1.33\, \rm{Mpc}/h$, gas temperature drops slightly from the boundary to around half of the radius, and then increases gradually toward the central region. For filaments with radius of $1.33-3.0\, \rm{Mpc}/h$, the temperature rises gently from outer region inward in the whole radial range and the temperature at center is higher than that at boundary by a factor of less than one. For thick filaments with $R_{fil}>3.0\, \rm{Mpc}/h$, the increase of average temperature in the outer regime of $r>0.8 R_{fila}$ and the inner regime $r<0.5 R_{fila}$ are also quite slow. In the middle regime $0.5<r/R_{fila}<0.8$, however, the temperature grows more rapidly, leading to the temperature at center is 2-3 times of that at the boundary. This rapid rise may have been caused by accretion shocks driven by gravitational collapse. Prominent filaments are more massive and likely to drive shocks. 

We find that the halos have minor influences on the volume weighted gas temperature profiles in filaments. The reason is that halos only occupy around $0.7\%$ of the volume in filaments at $z=0$. On the other hand, halos contribute $\sim 15\%$ of the mass in filaments. Comparing the left and right panels in the bottom row of Fig.~\ref{fig:fila_temprof}, we can see that the mass weighted temperature of filaments  $R_{fil}>2.0\, \rm{Mpc}/h$ would drop by a factor up to 2 if the contribution from halo gas is not taken into account. However, whether including halos or not has modest impact for the mass weighted temperature in filaments with  $R_{fil}<2.0\, \rm{Mpc}/h$. It is reasonable, because thicker filaments host more halos, especially those massive halos containing hot gases.   

\subsection{Evolution of filament profiles}
\begin{figure*}[htbp]
\begin{center}
\hspace{-0.0cm}
\includegraphics[width=0.45\textwidth, trim=0 10 10 10, clip]{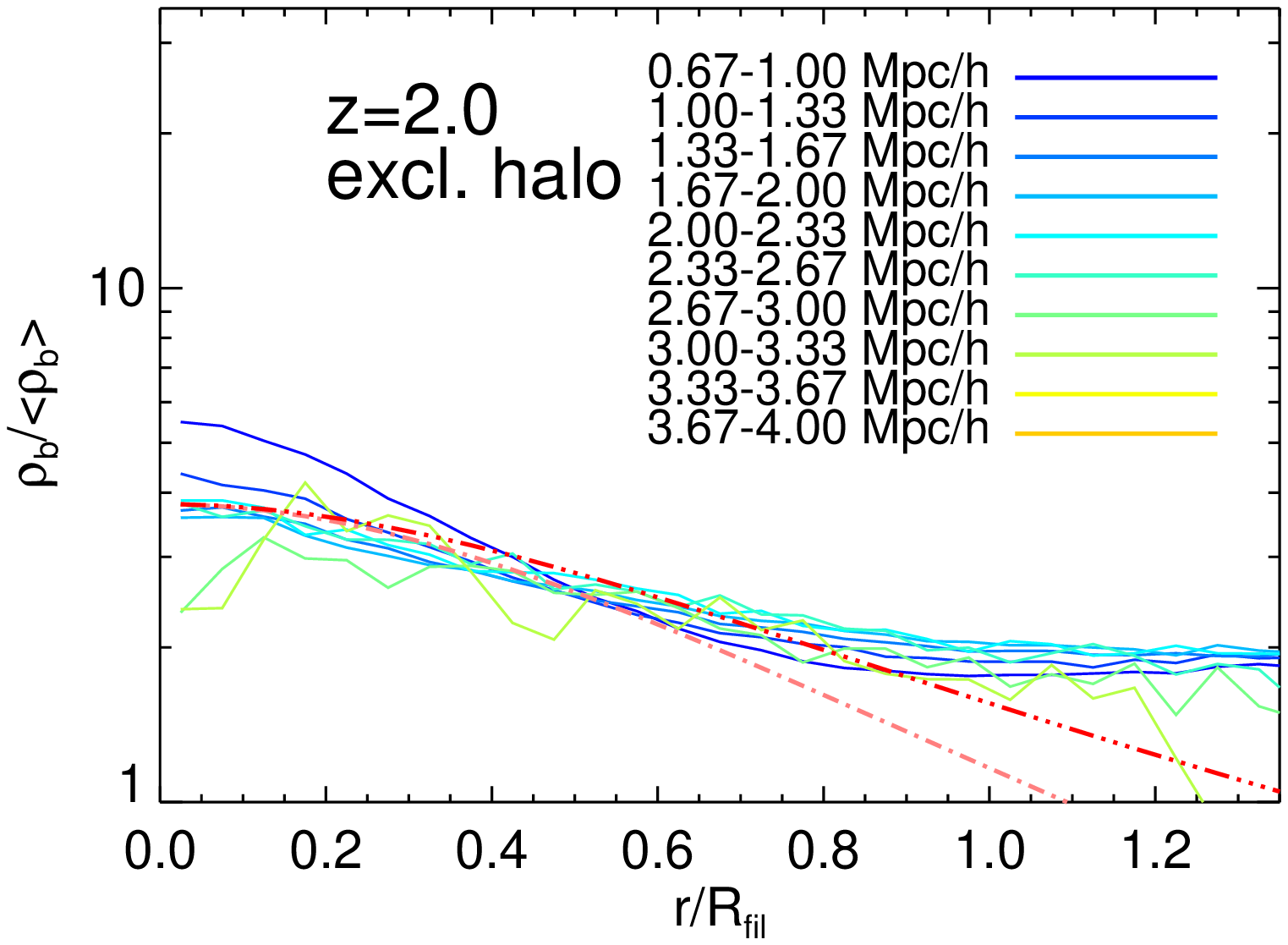}
\includegraphics[width=0.45\textwidth, trim=0 10 10 10, clip]{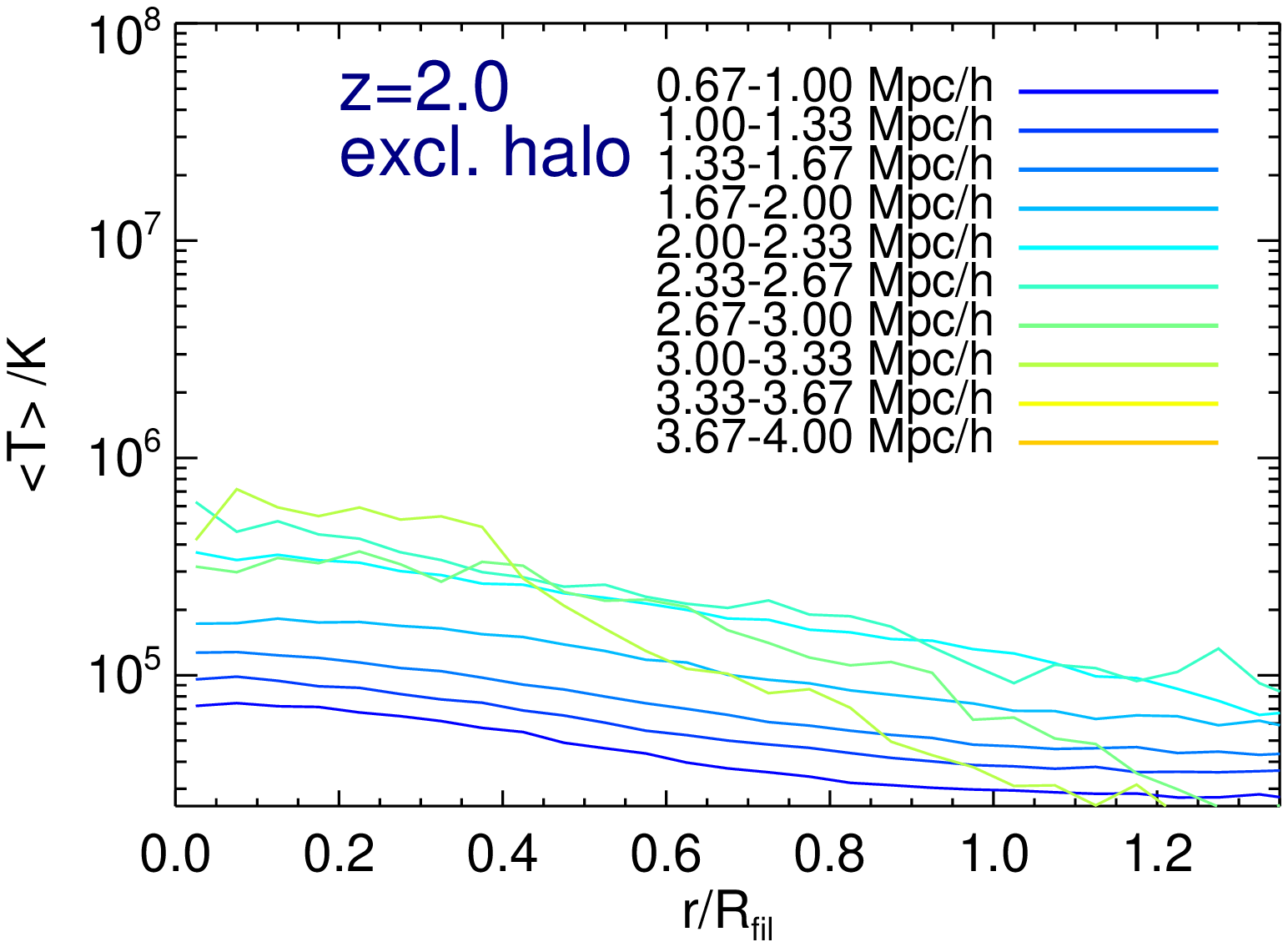}
\includegraphics[width=0.45\textwidth, trim=0 10 10 10, clip]{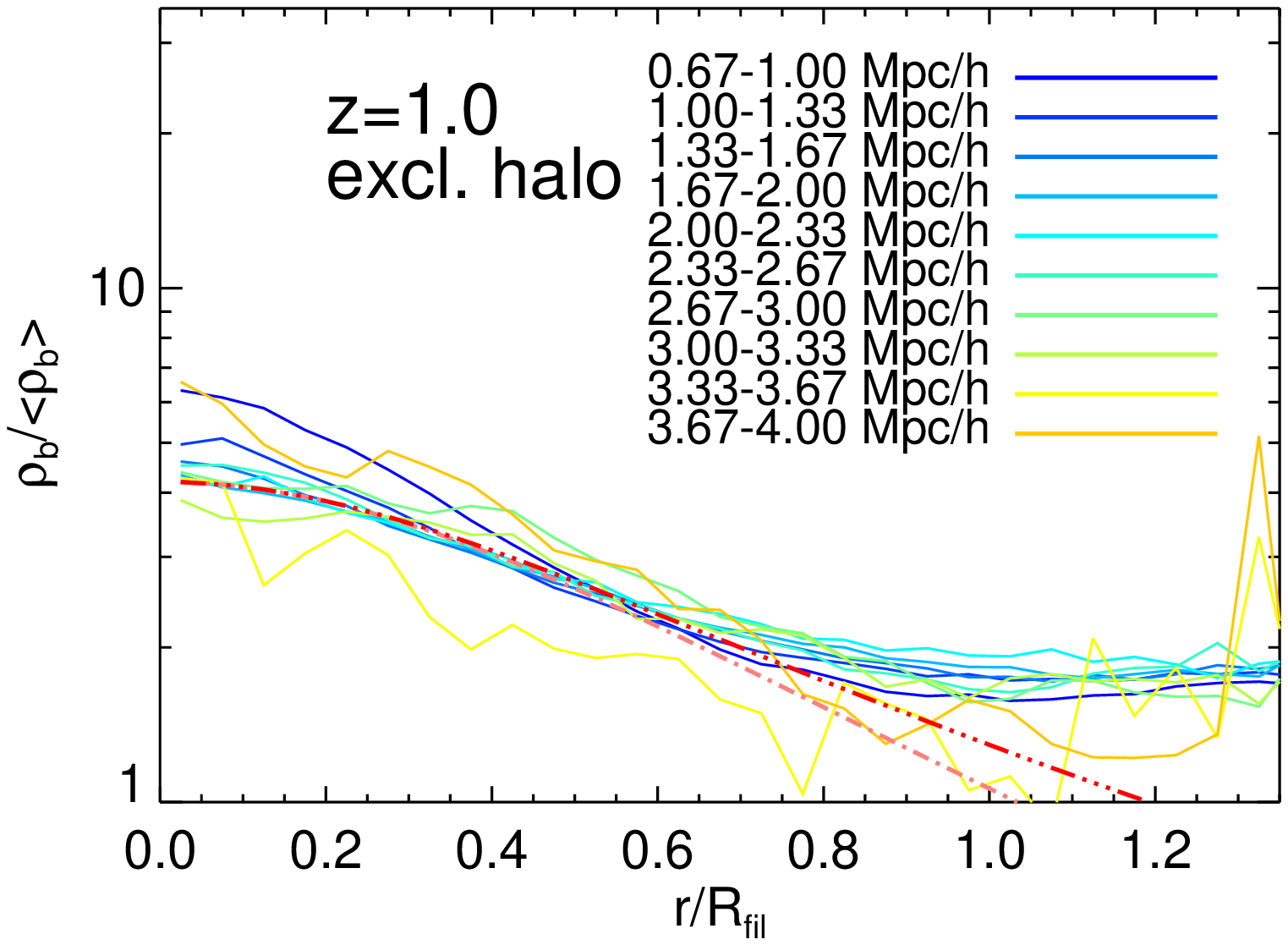}
\includegraphics[width=0.45\textwidth, trim=0 10 10 10, clip]{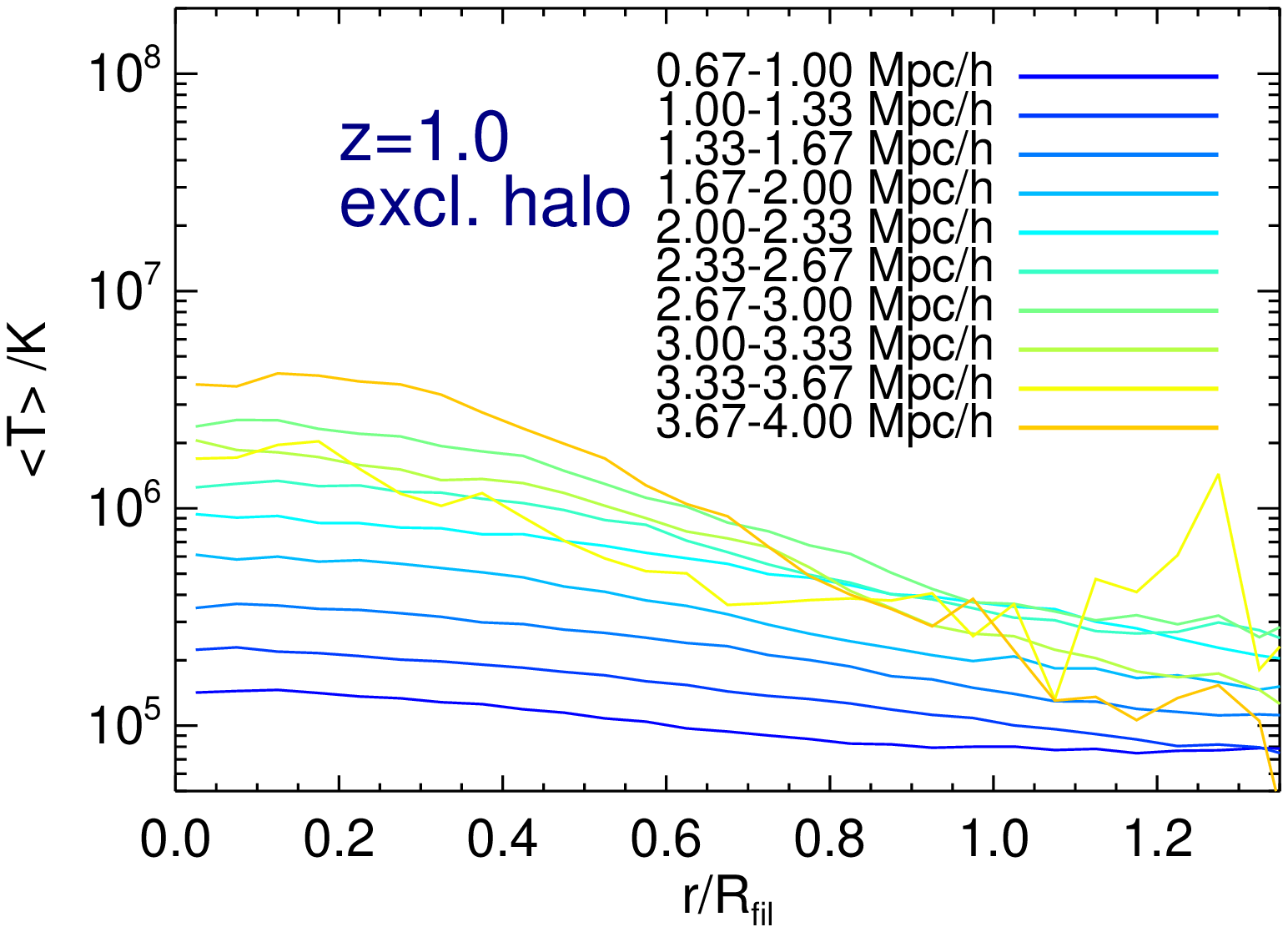}
\includegraphics[width=0.45\textwidth, trim=0 10 10 10, clip]{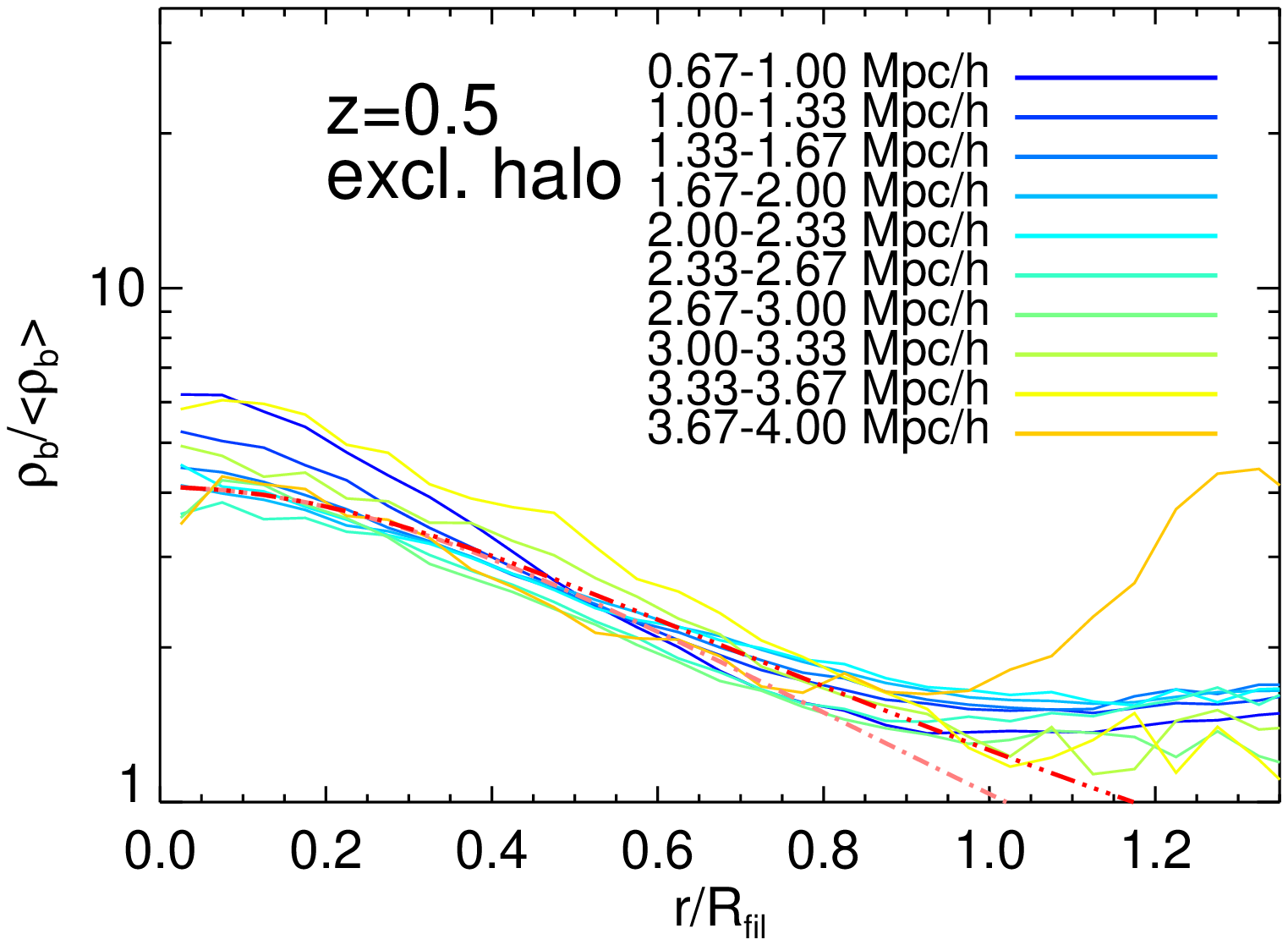}
\includegraphics[width=0.45\textwidth, trim=0 10 10 10, clip]{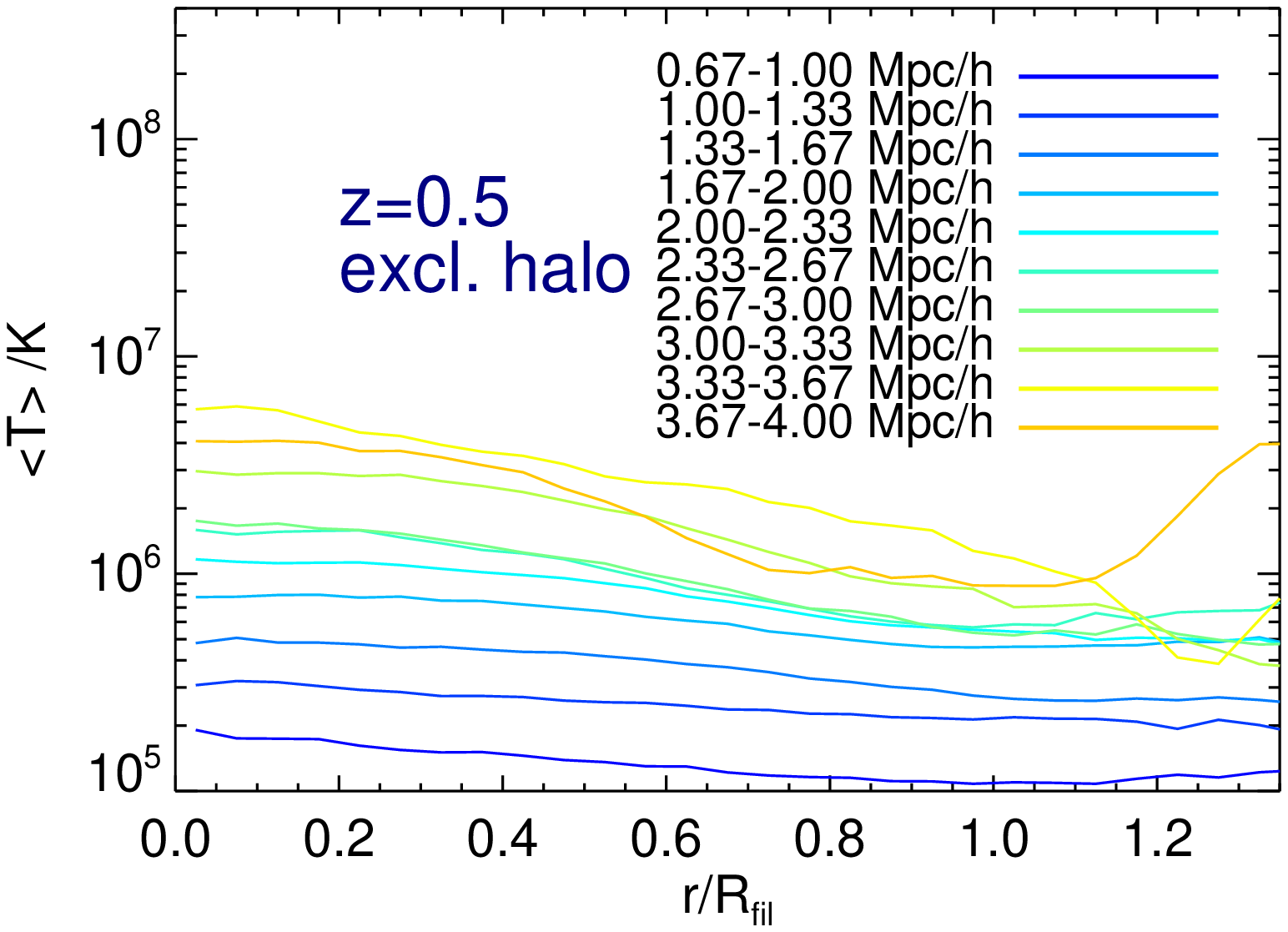}
\caption{Left: The density profile of baryon gas as a function of distance to the spine of filament, without the contribution from halos at redshift 2.0(top), 1.0(middle) and 0.5(bottom). Right: Same as the left column, but for the volume weighted temperature profile.}
\label{fig:fila_profevo}
\end{center}
\end{figure*}

In section 3.1, we show that the number frequency of tenuous filaments decreases slightly as redshift decreasing, which agrees with \citealt{2014MNRAS.441.2923C}. However, our work finds that the frequency of prominent filaments is increasing rapidly at $z<2$. Along with the number frequency, the profiles of filaments may also evolve with time. Figure \ref{fig:fila_profevo} shows the evolution of density and temperature profile of baryonic gas in filaments since $z=2.0$. For the results presented in Figure \ref{fig:fila_profevo}, the contributions from halos are excluded.

The density profiles at high redshifts can also approximately described by the isothermal single-beta model. Table ~\ref{tab:den_beta} lists the parameters that are used to fit the density profiles of gas and dark matter at different redshifts, with or without counting in the contribution from halos. The density peaks in the inner regime change slightly between $z=0.0$ and $z=1.0$, and decline moderately at $z>1.0$. Meanwhile, the density at radius $R_{fil}$ increases modestly with redshift increasing, which may result from a higher mean density in the overlapping regions of filaments and walls at high redshifts. Consequently, the density slope in filaments becomes more shallow at higher redshifts, leading to an extended core radius, i.e. larger $r_c$, in the single-beta formula. There are notable fluctuations in the profiles of thick filaments at high redshifts, due to the rapid decline of their number frequency as redshift increasing.

The temperature profiles of baryonic gas in filaments at high redshifts are similar to those at $z=0$. There are, however, two moderate differences. First, for filament with a given width, the overall gas temperature increases gradually with redshift decreasing. This feature is arising from the rise of mean temperature over time for gases in overdense regions due to UV background and gravitational collapse heating(e.g. \citealt{2017ApJ...847...17Z},\citealt{2019MNRAS.486.3766M}). Second, as redshift increases, baryonic gas hotter than $T=10^6 $K becomes more and more scarce in the cosmic filamentary structures, because there are fewer and fewer prominent filaments.

\begin{deluxetable}{cccc}[htbp]
\tablenum{1}
\tablecaption{parameters of the isothermal single-$\beta$ model used to fit density profile of gas and dark matter in filaments, where $\beta=1.0$}
\label{tab:den_beta}
\tablewidth{0pt}
\tablehead{
\colhead{Component} &  \colhead{$z$} & \colhead{$\rho_{0}/\bar{\rho}$} &\colhead{$r_c/R_{fil}$} 
}
\startdata
{ }& 0.0 & 6.5 & 0.8 \\
{ }& 0.5 & 6.0 & 0.8 \\
Gas(halo incl.) & 1.0 & 5.7 & 0.8\\
{ }& 2.0 & 4.4 & 0.9 \\
{ }& 3.0 & 3.6 & 0.9 \\
\hline
{ }& 0.0 & 4.1 & 0.8 \\
{ }& 0.5 & 4.1 & 0.8 \\
Gas(halo excl.)& 1.0 & 4.2 & 0.8\\
{ }& 2.0 & 3.8 & 0.9 \\
{ }& 3.0 & 3.6 & 0.9 \\
\hline
{ }& 0.0 & 6.8 & 0.8 \\
{ }& 0.5 & 6.8 & 0.8 \\
CDM(halo incl.)& 1.0 & 6.8 & 0.8\\
{ }& 2.0 & 4.8 & 0.9 \\
{ }& 3.0 & 3.8 & 1.0 \\
\hline
{ }& 0.0 & 4.3 & 0.8 \\
{ }& 0.5 & 4.8 & 0.8 \\
CDM(halo excl.)& 1.0 & 4.8 & 0.8\\
{ }& 2.0 & 4.4 & 0.9 \\
{ }& 3.0 & 3.9 & 0.9 \\
\enddata
\end{deluxetable}

\section{Discussions}
\label{sec:discussions}
\subsection{Impact of web classification}
\begin{figure}[htbp]
\begin{center}
\hspace{-0.0cm}
\includegraphics[width=0.42\textwidth, trim=0 285 10 10, clip]{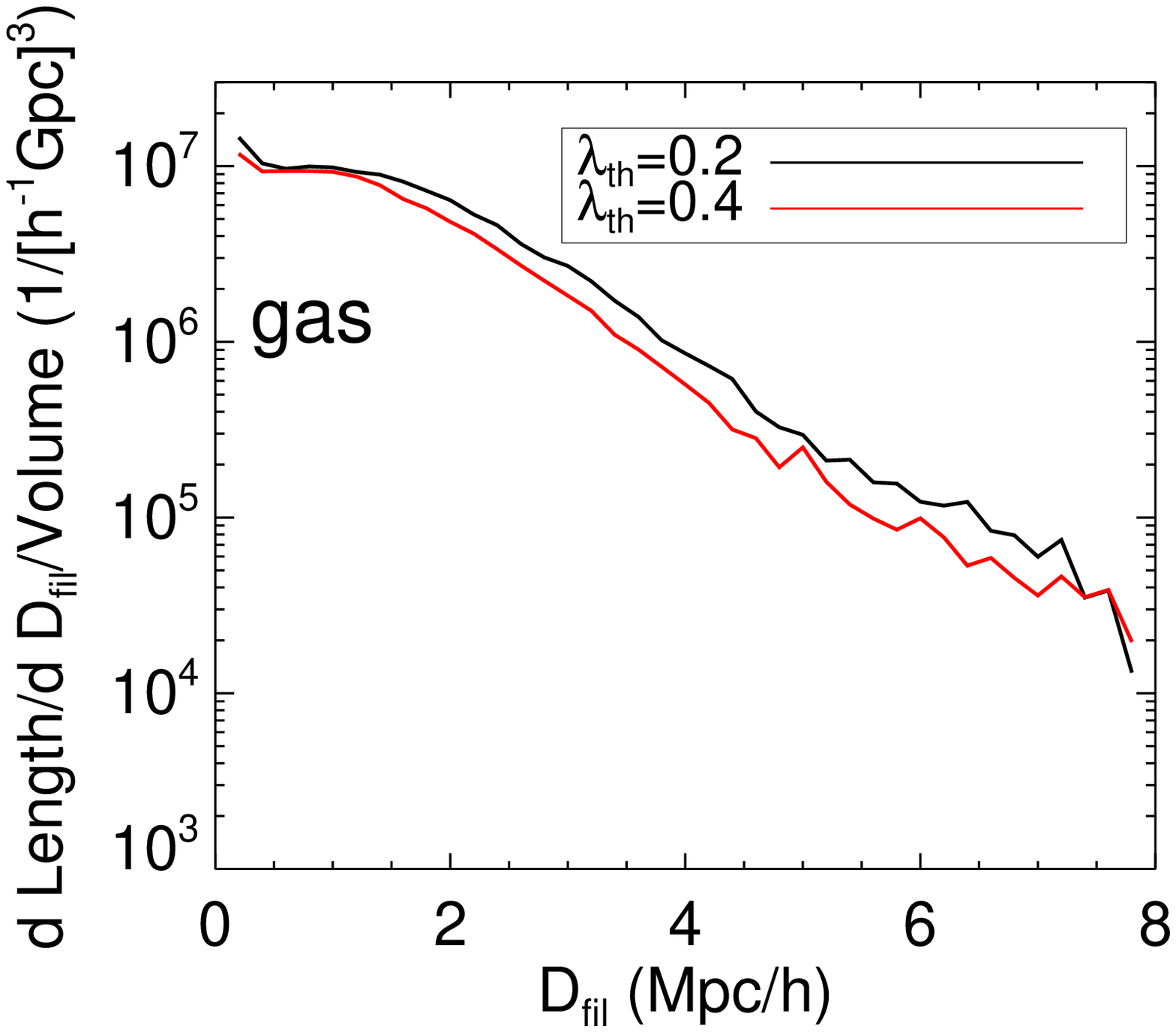}
\includegraphics[width=0.42\textwidth, trim=0 15 10 10, clip]{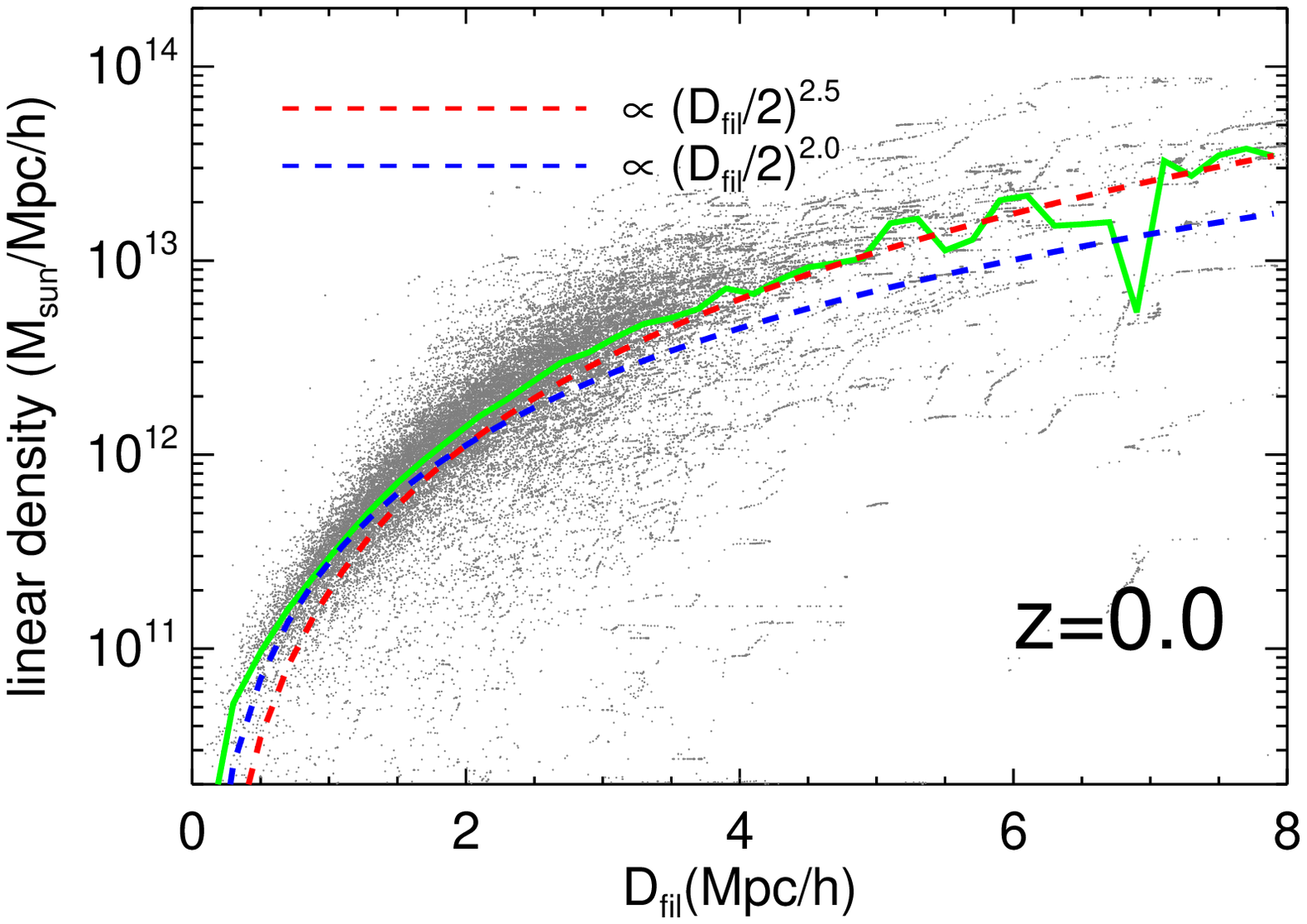}
\includegraphics[width=0.42\textwidth, trim=0 15 10 10, clip]{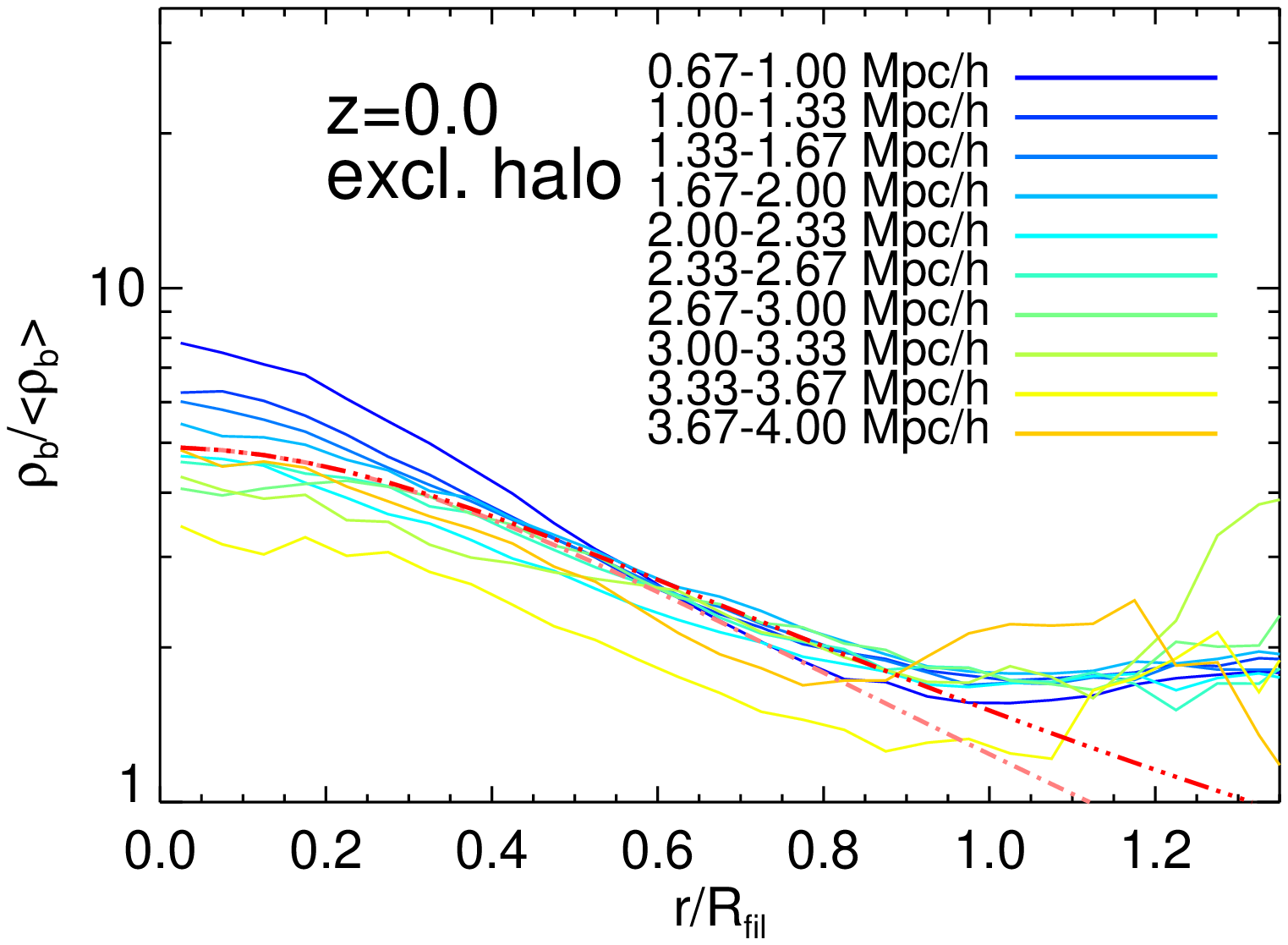}
\includegraphics[width=0.42\textwidth, trim=0 15 10 10, clip]{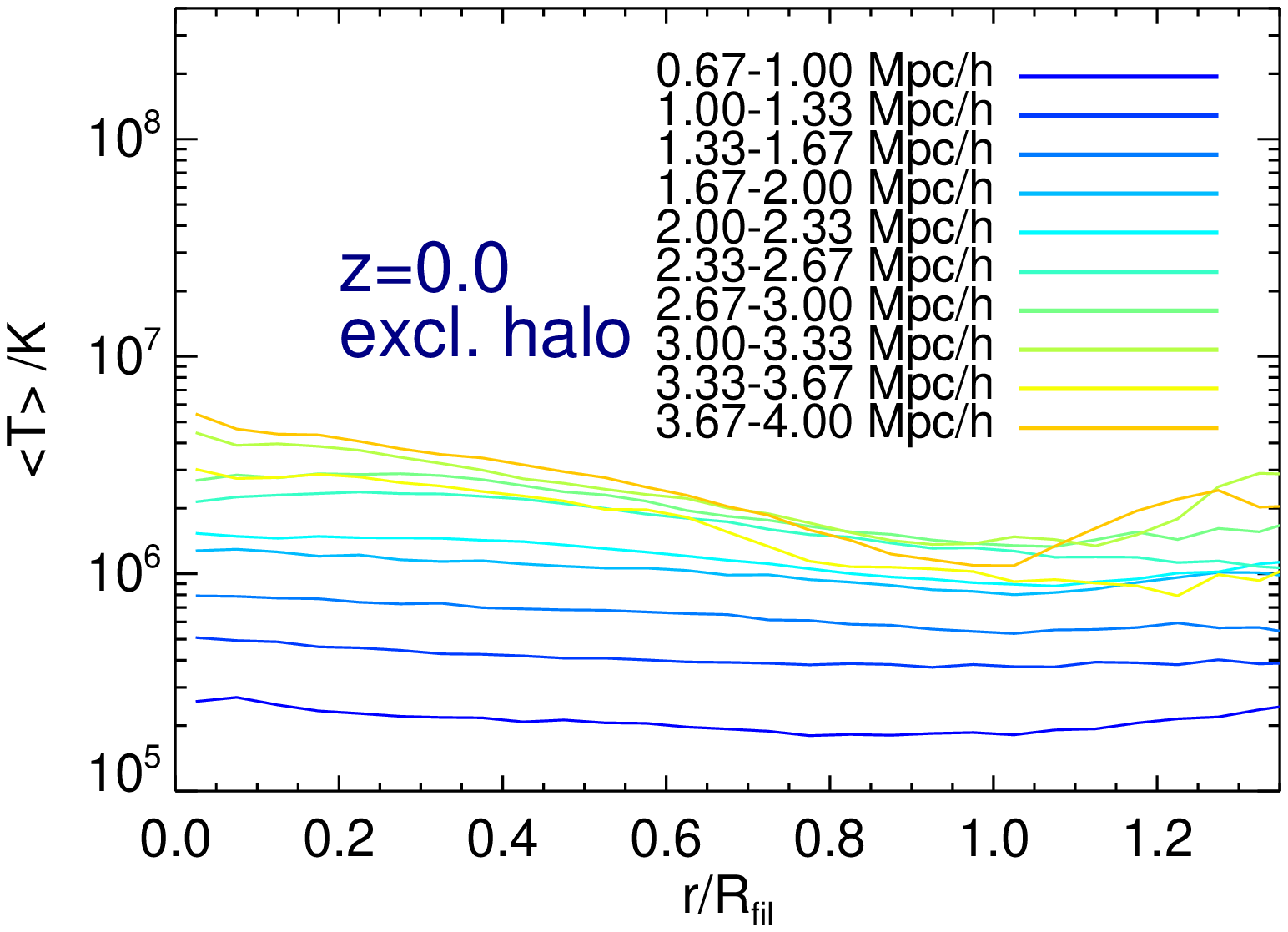}
\caption{Properties of filaments identified with $\lambda_{th}=0.4$. From top to bottom: the frequency as a function of local diameter; the scatter diagram of local diameter against local linear density; the density and volume weighted temperature profile of baryonic gas as a function of the normalised distance to the spine of filament, excluding the contribution from collapsed halos. }
\label{fig:fila4_prof}
\end{center}
\end{figure}

In the procedure of web types classification, the threshold eigenvalue $\lambda_{th}$ is an important parameter. Results presented in previous sections are based on a threshold eigenvalue of $\lambda_{th}=0.2$. It is worthwhile to examine whether these results are sensitive to the adopted value of $\lambda_{th}$. So far, a precise value of $\lambda_{th}$ that derived from the anisotropic collapse of structures is not available. In practice, $\lambda_{th}$ is taken currently ranging from $0.2$ to $0.6$ in relevant works involving the classification of large scale cosmic web(e.g. \citealt{2009MNRAS.396.1815F} \citealt{2017ApJ...838...21Z}; \citealt{2019MNRAS.486.3766M}).
We perform a similar analysis as the previous sections for the simulation sample at $z=0$ but with $\lambda_{th}=0.4$, and show the results at $z=0$ in Fig.~\ref{fig:fila4_prof}. 

A larger $\lambda_{th}$ leads to less grid cells residing in filaments of almost all width, which is consistent with our expectation. On the other hand, the correlation between the linear density and the local diameter persists, although the linear density coefficient in the right hand side of Eqn.~\ref{eqn:fila_linden} increases $\sim 20\%$. The density profiles in filaments with different widths are still similar, and can be approximately described by a single-beta model with the same value of $r_c$ and $\beta$ as the case of $\lambda_{th}=0.2$. Yet, the overall densities have been increased by $\sim 20\%$ from the inner to the boundary regions with respect to $\lambda_{th}=0.2$. 

The shape of temperature profiles shows minor changes when $\lambda_{th}$ goes from 0.2 to 0.4, but the typical temperature for filaments with a particular width rise modestly. The reason is that the outer boundary of filaments would move inward to the spine of filaments or to the center of nodes/clusters as $\lambda_{th}$ increases. Consequently, for filaments with the same width but classified with different thresholds $\lambda_{th}$, the typical density and temperature would be systematically higher for larger $\lambda_{th}$. 

\subsection{Comparison to previous simulation works}
The density and temperature profiles of baryonic gas in filaments have been studied in recent works such as \cite{2019MNRAS.486..981G}, \cite{2020arXiv201015139G} and \cite{2020arXiv201209203T}. In \cite{2019MNRAS.486..981G}, the properties of filaments (e.g. density, temperature,velocity, magnetic field) across a few orders of magnitude in mass have been investigated. In \cite{2020arXiv201015139G}, filaments in three different length ranges are averaged respectively to calculate the corresponding profiles.  \cite{2020arXiv201209203T} takes an average over all the found filaments, and a sub-sample of the filaments respectively to obtain the profiles. A comparison with these works could provide a more comprehensive view on the properties of filaments. We first take a look to the temperature profiles, and then to the density profiles. 

The temperature profiles show an isothermal core and then drops either sharply or gradually at outer regions in all the three works mentioned above. We also find an isothermal core region in the profiles of our samples, continuing with declining outward to outer region, despite the differences in simulation samples, web classification and profile averaging methods. However, the radius of isothermal central part in our samples increases with the local filament width, which is different from the literature. For thick filament segments with radius $R_{fil}>2 \rm{Mpc}/h$, the isothermal core radius can extend to 1-2 Mpc in our work. This is comparable to the radius of isothermal core in \cite{2019MNRAS.486..981G} and \cite{2020arXiv201015139G}. Nevertheless, for filament segments with $R_{fil}<2 \rm{Mpc}/h$, the isothermal radius is smaller, and more close to the core radius in \cite{2020arXiv201209203T}.

The temperature peaks in the central core region of filaments varies from $10^5\, $K to $3\times 10^6 \, $K in the previous studies, which agree with the overall temperature range in our samples. Our analysis shows that the typical temperature of gas is hotter in thicker and more prominent filaments, which usually locate in more dense regions. This feature is generally consistent with \cite{2020arXiv201015139G} and \cite{2020arXiv201209203T}. \cite{2020arXiv201015139G} shows that the temperature of short filaments in dense regions is about $1.3 \times 10^6$K, i.e., 3 times of long filaments that traces less-dense regions. Moreover, shorter filaments in \cite{2020arXiv201015139G} tend to be thicker.  \cite{2020arXiv201209203T} find that the filaments with high luminosity density (tend to have high overdensity), have a peak temperature $\approx 1.2\times 10^6$ K, which is 12 times of the full filaments sample. All these works are consistent with the picture that prominent and thick filaments, usually form in more dense region and more close to nodes/clusters, are much hot than than those thin filaments. 

In our work, only grid cells in the environment of filaments are included to calculate the profiles of filaments. Therefore the temperature profiles are truncated at a radius moderately larger than $R_{fil}$. In contrast, the average temperature at radius far away from the outer boundary is available in \cite{2020arXiv201015139G} and \cite{2020arXiv201209203T}, and is about $10^4-10^5 \, $K, which is close to the average background temperature of the IGM. 

The density profiles of filaments in \cite{2019MNRAS.486..981G} and \cite{2020arXiv201209203T} have been presented explicitly, while it is not shown directly in \cite{2020arXiv201015139G}, and so it is somehow difficult to make a detail comparison with \cite{2020arXiv201015139G}. The density profile presented in \cite{2019MNRAS.486..981G} for a given filament is similar to our results by visual check.  \cite{2020arXiv201209203T} used an isothermal single-beta model with $\beta=2.0$ to fit the density profiles of filaments with high luminosity density, results in a gas density peak around $20$ times of the cosmic mean and a core radius $r_{c,\rho}=1.2 Mpc$, i.e, a profile steeper than our samples. 

The overall values of density in our samples are close to \cite{2019MNRAS.486..981G}, which has a peak about 5-10 times of the cosmic mean, $<\rho_b>$, at $r=0$ and drops to $\sim 3$ at $r>2 Mpc$. For the full samples of filaments in \cite{2020arXiv201209203T}, the baryon overdensity at filament spine is about $\rho_b/<\rho_b>\approx5$, in agreement with our results. Nevertheless, for their filaments with high luminosity density, the baryon overdensity at center is much higher, up to $\rho_b/<\rho_b>\approx40$. This discrepancy may partly because we only sum over grids in the environment of filaments to calculate the average density, and discard those grids in the environment of nodes/clusters. Therefore, some grid cells outside of the halos in high overdensity regions are removed. In \cite{2020arXiv201209203T}, only gas cells within Friend-of-Friend halos are excluded. Moreover, the EAGLE simulation used in \cite{2020arXiv201209203T} has a higher resolution with respect to our simulation, which may enhance the density in the central regime of filaments. 

We conclude that, the ranges of the density and temperature in the filaments of our simulation are generally consistent with the literature. Nevertheless, there are notable discrepancies in the shape of profiles, core radius in comparison with many works, and even in the gas overdensity in the central core regions with some works. These differences should partly result from the web classification procedures, and the average methods in calculating the profiles as well as different sample sizes. In addition, the differences in the simulation schemes, resolution and sub-grid physics may also have contributed to the diversity.

\subsection{Comparison to detection of WHIM in filaments via tSZ signals}
Detection of the thermal SZ signal due to the warm and hot gas in filaments have been reported recently by several groups(\citealt{2019MNRAS.483..223T};\ \citealt{2019A&A...624A..48D}; \ \citealt{2020A&A...637A..41T}).
The former two works searched for signals caused by filaments between galaxy pairs taken from the Sloan Digital Sky Survey(SDSS), while the latter probes signals due to gas in filaments with lengths of 30 to 100 Mpc identified in SDSS survey.   \cite{2019MNRAS.483..223T} found a Compton y-parameter $y \approx 1 \times 10^{-8}$ for filaments between pairs of Luminous Red Galaxies at redshifts $z<0.4$. Assuming the matter in filament follows an isothermal single-beta model with $\beta=2/3$, they obtain an estimation on the product of over-density and temperature to be $\delta_c\times(T_e/10^7\rm{K})\times(r_c/0.5h^{-1}\rm{Mpc})=2.7\pm0.5$. This is in good agreement with our results for filaments with $R_{fil}>2.0$ Mpc, which have a over-density $\sim 4-5$(halos excluded), temperature around $1.5-4\times 10^6$K and core radius $0.8\times R_{fil}$. \

\cite{2019A&A...624A..48D} reported a mean Compton parameter $y \approx 0.6\pm \times 10^{-8}$ due to WHIM by stacking one million pairs of CMASS galaxies in the redshift range $0.43<z<0.75$(mean 0.55) from SDSS survey. Assuming the gases in filaments follow a Gaussian profile with a FWHM of $1.5h^{-1}$ Mpc along the axis perpendicular to the spine of filaments, they estimated that a central gas density about $5.5\pm2.9$ times of the cosmic mean and a gas temperature $\sim 2.7\times 10^6$ K can explain the tSZ signal they detected. These values are also consistent with the profiles of thick filaments at $z=0.5$ in our sample, although the shape of profile is different.   

\cite{2020A&A...637A..41T} detected the tSZ signal of filamentary structures at a significance of $4.4 \sigma$. The central overdensity in filaments are estimated to be around $19.0_{-12.1}^{+27.3}$ and $6.3_{-0.8}^{+0.9}$ for isothermal $\beta$ model($\beta=2/3$) and constant density model respectively. The former value is higher than our result. The electron temperatures are found to be $\sim 1.3-1.4\times 10^6$ K for two models, which is consistent with the values in our samples. 

In short, the gas over-density and temperature of thick filaments in our sample agree with most of the corresponding estimation in the recent observational works that have reported the detection of tSZ signal attributed by filaments. Yet, we should remind the readers that the filament classification and averaging method we used are different with those used in observational works.

\section{Summary} 
\label{sec:summary}
A theoretical knowledge of the distribution of baryonic matter in filaments is desired for the search for missing baryon. In this work, We have studied the properties of cosmic filaments since $z=4.0$ in a cosmological hydrodynamic simulation with adaptive mesh refinement. We have quantitatively evaluated the evolution of filaments with different width after $z=4$, and measured the density and temperature profiles perpendicular to the filament spine. Our findings are summarized as follows:
\begin{enumerate}
\item Quantitative evaluation shows that the number frequency of thick filaments grows rapidly after $z=2$, in good agreement with the visual impression presented in the literature. Filaments with width $D \gtrsim 4.0 \, \rm{Mpc}/h$ comprise $\sim 10\%$ of the baryon contained in filaments at $z=2$, i.e., $\sim 4\%$ of the baryon in universe. These fractions grow up to $\sim 28\%$ and $\sim 13\%$ respectively at $z=0$. 
\item The linear density of filaments, i.e., the mass contained in a unit length along the filament spine, is correlated with the local diameter, and approximates a power law function $\propto (D_{fil})^n$, since redshift as high as $z=4$. The power index increases gradually from $n\approx 2$ at $z=4$ to $n\approx2.5$ at $z=0$.
\item The averaged density profiles of both dark matter and baryonic gas in filaments of different widths show self-similarity, and can be described by an isothermal single-beta model at redshift $z<4$, as Eqn. \ref{eqn:den_cdm} and \ref{eqn:den_gas}. The profiles for the baryonic gas and dark matter are quite similar. The density profiles become more shallow with increasing redshifts.
\item The gas temperature profiles in thin filaments are relatively flat in the whole radial region, while the temperature rises significantly in the middle redial regime for filaments with local width $D_{fil} \gtrsim 4.0 \rm{Mpc}$. The typical gas temperature increases with the filament width increasing, and is hotter than $10^6$ K for filaments with width $D_{fil} \gtrsim 4.0 \rm{Mpc}$, and vice versa.  Filament segments with $D_{fil}> 4.0 \rm{Mpc}$ would dominate the tSZ signal caused by baryons in filaments. 
\end{enumerate}
Further investigation with other simulations samples is urged to verify the results revealed by our work. On the other hand, to make our results more conducive to indirect search for the missing baryon, it would be necessary to optimise the strategy of constructing filament samples based on observation data. These studies will be carried out in the future. 

\acknowledgments
We thank the anonymous referee for her/his useful comments to improve the manuscript. This work is supported by the National Natural Science Foundation of China (NFSC) through grant 11733010. W.S.Z. is supported by NSFC grant 11673077. Z.F.P. is supported by 2021A1515012373 from Natural Science Foundation of Guangdong Province. The cosmological hydrodynamic simulation was performed on the Tianhe-II supercomputer. Post simulation analysis carried in this work was completed on the HPC facility of the School of Physics and Astronomy, Sun, Yat-Sen University.
%

\vspace{5mm}


\software{RAMSES \citep{2002A&A...385..337T},  
          }





\bibliography{main}{}
\bibliographystyle{aasjournal}



\end{document}